\documentclass[12pt]{article}
\usepackage{epsfig,amsfonts,amssymb}
\usepackage{amsmath,graphics}

\usepackage{hyperref}
\usepackage{cleveref}
\crefformat{footnote}{#2\footnotemark[#1]#3}
\def\ZZZ{{\hbox{ Z\kern-1.6mm Z}}}
\def\RRR{{\hbox{ R\kern-2.4mm R}}}
\def\CCC{{\hbox{ C\kern-2.0mm C}}}
\def\zzz{{\hbox{z\kern-1mm z}}}
\newcommand{\mathsym}[1]{{}}

\newcommand{\qeq}{{\hbox{=\kern-2.3mm ? \kern.5mm }}}
\renewcommand{\qeq}{=}

\newcommand{\be}{\begin{equation}}
\newcommand{\ee}{\end{equation}}
\newcommand{\ben}{\begin{eqnarray}\displaystyle}
\newcommand{\een}{\end{eqnarray}}

\def\one{{\hbox{ 1\kern-.8mm l}}}
\def\zero{{\hbox{ 0\kern-1.5mm 0}}}

\begin{document}

\baselineskip 24pt

\begin{center}
{\Large \bf On the smoothness of horizons in the most generic  multi center black hole and membrane solutions.}

\end{center}

\vskip .6cm
\medskip

\vspace*{4.0ex}

\baselineskip=18pt

\centerline{\large \rm   Chethan N. Gowdigere}

\vspace*{4.0ex}

\centerline{\large \it National Institute of Science Education and Research.}

\centerline{\large \it  Sachivalaya Marg, PO: Sainik School,}

\centerline{\large \it  Bhubaneswar 751005, INDIA}

\vspace*{1.0ex}
\centerline{E-mail: chethan.gowdigere@niser.ac.in }

\vspace*{5.0ex}

\centerline{\bf Abstract} \bigskip
We study the differentiability of the metric and other fields at any of the horizons  of  the \emph{most generic} multi center Reissner-Nordstrom black hole solutions in $d \ge 5$ and of multi center $M2$ brane solutions. 
Most generic means that the centers are generically located in transverse space and consequently the solutions do not have any transverse spatial  isometries.  We construct the Gaussian null co-ordinate system for the neighborhood of a  horizon by solving (all) the geodesic equations in expansions of (appropriate powers of) the affine parameter. Organizing the harmonic functions that appear in the solution in terms of generalized Gegenbauer polynomials, introduced in \cite{Gowdigere:2014aca}, is key to obtaining the solution to the geodesic equations in a compact and manageable form.  We then compute the metric and other fields in the Gaussian null co-ordinate system and find that the differentiability of the horizon in the most generic solution  is \emph{identical to} the differentiability of the horizon in the two center/collinear solution (centers distributed on a line in transverse space).  We isolate those aspects of the computation that are most relevant to this result. We perform these computations in some cases, in  several  co-ordinate systems.

\vfill \eject

\baselineskip=18pt

\tableofcontents

\section{\label{1}Introduction}
In this paper, we continue with and bring to an end (one aspect of) the  study of smoothness/differntiability of horizons   in $d \geq 5$ Reissner-Nordstrom multi center black holes and in multi membrane solutions in M-theory; previous studies are contained in \cite{Hartle}, \cite{Gibbons}, \cite{Welch}, \cite{Candlish1}, \cite{Candlish2}, \cite{Gowdigere:2012kq}, \cite{Gowdigere:2014aca}.
By studying horizon smoothness/differentiability is meant the determining of the degree of differentiability/smoothness (smooth being $\mathcal{C}^\infty$, only $k$-times differentiable $\mathcal{C}^k$) at the horizon of the (components of the) various fields present in the solution such as the metric, gauge fields, tensor gauge fields. A horizon is smooth if all components of all tensor fields of the solution are smooth at the horizon; else one says that the horizon is not smooth and further supplements the statement by giving the degree of smoothness/differentiability of the various tensor fields of the solution; the degree of smoothness/differentiability of a tensor field being the smallest of the degrees of differentiability of all the components of the tensor field.

Equations \eqref{rnsoln}, \eqref{rnsoln2} contain the black hole solutions  and  equations \eqref{m2soln}, \eqref{m2harm} the multi-membrane solutions that we study, given in the isotropic co-ordinates.  Both classes of solutions have a common feature:  a part of the space-time is conformally a Euclidean space,  conformal $\mathbf{R}^{d-1}$ in the black hole case and a conformal $\mathbf{R}^{8}$ in the membrane case, often referred to as the transverse Euclidean space. Furthermore each of these solutions is completely specified by an arbitrary harmonic function, the $H$ that appears in the equations \eqref{rnsoln2} and \eqref{m2harm}, harmonic in the transverse Euclidean space. When $H = 1 + \frac{\mu}{r^{d-3}}$ in the black hole case and when $H = 1 + \frac{\mu}{r^6}$ in the membrane case, the solutions describe a single black hole and a single membrane respectively; these solutions are referred to as  single center solutions. Apart from the $\frac{\partial}{\partial t}$ static isometry in the black hole case and the $\frac{\partial}{\partial t}, \frac{\partial}{\partial x}, \frac{\partial}{\partial y}$ brane translation isometries in the membrane case, there are spatial rotational isometries: $\mathbf{so}(d-1)$ for the black holes  and a $\mathbf{so}(8)$ in the membrane case. The only horizon of the single center solutions is known to be smooth.  When the harmonic function has two centers,  the spatial rotational symmetries of the solution are only those rotations in the transverse Euclidean space that preserve the line joining the two centers. They constitute a $\mathbf{so}(d-2)$ in the black hole case and a $\mathbf{so}(7)$ in the membrane case. This is true even for an arbitrary number of centers all located on one line.  Still, one refers to this as the ``two center'' case, two being the number of centers in generic positions (the others are not in generic positions but can only be positioned on the line joining the first two) and sometimes also as the ``collinear'' case.   The analysis of the smoothness of horizons in two center solutions, for the $d \geq 5$ black hole case, was done in \cite{Candlish1} by Candlish and Reall (building on earlier work by \cite{Welch}), where it was found that the horizons are not smooth. At any of the horizons, for $d=5$, the metric was found to be only $\mathcal{C}^2$ and the gauge field was found to be $\mathcal{C}^0$ (continuous and not differentiable) and for $d \ge 6$ the metric was found to be only $\mathcal{C}^1$ and the gauge field $\mathcal{C}^0$. The two center membrane solutions were analyzed by some of us in \cite{Gowdigere:2012kq} (the first indication that they may not be smooth was there in \cite{Gibbons}), where it was found that horizons are not smooth: the metric was found to be only $\mathcal{C}^3$ while the tensor gauge field  was only $\mathcal{C}^2$ at any of the horizons.

Going on, when the harmonic function has three centers in generic positions (or even an arbitrary number - greater than three - of centers all distributed on a plane) in the transverse Euclidean space, the spatial rotational symmetries of the solution constitute  a $\mathbf{so}(d-3)$ in the black hole case and a $\mathbf{so}(6)$ in the membrane case. One  refers to this situation as the ``three center'' case and also as the  ``coplanar'' case. The analysis of the smoothness of horizons in three center solutions for both  black holes and membranes was done by us in \cite{Gowdigere:2014aca}, where it was found that none of the horizons involved are smooth. Moreover, the degree of smoothness of the horizon in the three center solutions was found to be exactly \emph{identical} to the degree of smoothness of the horizon in the two center solutions.

The procedure to obtain these reults was laid out in \cite{Candlish1} and essentially repeated in \cite{Gowdigere:2012kq} and \cite{Gowdigere:2014aca} except with some modifications to allow for the peculiarities of the membrane horizon. We will describe this procedure, even here in the introduction, partly because it is needed to describe the setting for the present work and also because it is the procedure we follow here. We will describe the procedure mainly for the black hole case here.  The goal is to study the smoothness properties of various tensor fields at the horizon of the (first) black hole. The solutions \eqref{rnsoln} \eqref{m2soln} are presented in the isotropic co-ordinate system: for the black hole isotropic co-ordinates are the $t$ that appears in \eqref{rnsoln} and any co-ordinate system for the transverse Euclidean space which comprises a radial co-ordinate $r$ and $d-2$ angles $\theta_1, \theta_2, \ldots \theta_{d-2}$;  for the membrane case isotropic co-ordinates are the $t, x, y$ that appears in \eqref{m2soln} and any co-ordinate system for the transverse Euclidean case.  But the isotropic co-ordinate patch  does not cover any of the horizons. Hence, one needs to first construct a good horizon co-ordinate system. Following \cite{Candlish1}  we work with a horizon co-ordinate system known as the Gaussian null co-ordinate system. We will not give the full theory of Gaussian null co-ordinates here. For this, we refer, apart from  the original reference \cite{Friedrich:1998wq}, to \cite{Candlish1} for a good summary (see also \cite{Gowdigere:2014aca}).  The Gaussian null co-ordinate system is constructed using radial null geodesics. One first obtains the solution to the geodesic equations $t(\lambda), r(\lambda), \theta_1(\lambda), \ldots \theta_{d-2}(\lambda)$. There are $d-1$ integration constants appearing in the solution: $v$, $\Theta_1, \Theta_2 \ldots \Theta_{d-2}$ (see section \ref{2} for details).  It turns out, from the theory, that the affine parameter $\lambda$ together with these integration constants comprise a good co-ordinate system for the horizon and its neighbourhood, referred to as the Gaussian null co-ordinates. The solution to the geodesic equations, now written as $t(\lambda, v, \Theta_1 \ldots \Theta_{d-2}), ~r(\lambda, v, \Theta_1 \ldots \Theta_{d-2}) \ldots$ are the transition functions between the isotropic and the Gaussian null co-ordinates.  Except for the single center case, it is hard to obtain the exact solution to the geodesic equation. One makes a series expansion ansatz, the expansion parameter an  appropriate power of the affine parameter\footnote{The fact that the correct expansion parameter is some fractional power of the affine parameter  rather than the affine parameter itself is the technical reason why  there is finite differentiability. For $d=4$ black holes, it turns out that the expansion parameter is nothing but the affine parameter and there is no finite differentiability around any of the multi center horizons, which is the result of \cite{Hartle}, although they use different methods.}  (see ahead \eqref{264}), for each of the unknown functions  $r(\lambda), \theta_1(\lambda), \ldots \theta_{d-2}(\lambda)$, and plugs them in to the geodesic equations, and solving order by order, obtains the coefficients in the series expansion.

Having thus obtained a good horizon co-ordinate system, one then proceeds to study smoothness as follows. The transition functions between the two co-ordinate patches are  used to obtain the component functions of the various tensor fields of the solution in the Gaussian null co-ordinate patch via the tensor transformation law. Since the transition functions are series expansions, the expressions for the component functions are also series expansions. By examining these series expansions for negative or fractional powers of the affine parameter, one reads off their degree of differntiability;  one would need to compute the series expansions till at least the first fractional power of the affine parameter. From the degrees of differentiability of all components of all tensor fields, one  obtains the answer for the smoothness of the horizon.

The above description of the technical details of determining horizon smoothness, allows us to describe in more detail the results of \cite{Gowdigere:2014aca}, beyond the statement that  the degree of smoothness of the horizon in the three center solutions is exactly \emph{identical} to the degree of smoothness of the horizon in the two center solutions. The harmonic function in the two center solution is a function of the radial co-ordinate $r$ and one\footnote{\label{second} in a certain \eqref{spcoordinates}, not every, choice of co-ordinates for the the transverse Euclidean space (see \ref{214} for a different choice when this does not hold).} of the angles $\theta_1$. The metric is diagonal and the gauge field has only one non-zero component $A_t$ \eqref{rnsoln}. In the Gaussian null co-ordinate system, the metric  has non-zero  off-diagonal components viz. $g_{\lambda v}, g_{v\Theta_1}$ besides the diagonal ones (except $g_{\lambda \lambda}$\footnote{\label{third}In fact, from the theory of Gaussian null co-ordinates, it follows that $g_{\lambda \lambda}= 0, ~g_{\lambda v} = 1, ~g_{\lambda \Theta_i} = 0$, see \cite{Gowdigere:2014aca}.}) and the gauge field has non-zero components $A_{\lambda}, A_v, A_{\Theta_1}$. The harmonic solution in the three center solution is a function of the radial co-ordinate $r$ and two\cref{second} of the angles $\theta_1, \theta_2$. In the Gaussian null co-ordinate patch, the three center metric and gauge field have additional non-zero components besides the ones which were non-zero for the two center situation viz. $g_{v\Theta_2}, g_{\Theta_1 \Theta_2}, A_{\Theta_2}$. Apart from the generic statement that the degree of smoothness of the horizon in the three center solution is identical to the degree of smoothness of the horizon in the two center solution, we also made some more precise observations \cite{Gowdigere:2014aca}: When going from the two center to the three center case, only one of the following three things happen for tensor components in the Gaussian null co-ordinate system:

\begin{itemize}
\item (\textbf{P1}) Components which were smooth in the two center solution continue to be  smooth in the three center solution i.e. in the series expansions there are no terms with fractional or negative powers of the affine parameter.

\item (\textbf{P2})  Components which were  smooth in the two center solution become non-smooth i.e.  there are non-zero coefficients for fractional or negative powers of the affine parameter in the series expansion for the component in the three center solution.  But the resulting finite degree of differentiability does not change the degree of smoothness of the tensor field and hence the horizon smoothness is unchanged.

\item (\textbf{P3})  Components which had a finite degree of smoothness in the two center solution are modified but the modifications preserve the series expansion, changing only the coefficients which were already non-zero. Thus there is no modification to the degree of differentiability of the component itself as well as the degree of smoothness of the tensor field and also of the horizon.
\end{itemize}
$g_{v\Theta_2},~g_{\Theta_1 \Theta_2}$ and $A_{\Theta_2}$ follow (\textbf{P2}), all components which were non-zero in the two center solution (except\cref{third} $g_{\lambda v}$)  follow (\textbf{P3}) and the rest (\textbf{P1}). 
Two other logically allowed possibilities, which don't seem to be realized in the results,  are as follows. One is the opposite of (\textbf{P2}) i.e.  that  components acquire a degree of differentiability less than the the degree of differntiability of the tensor field in the two center solution,  which would result in the horizon of the three center  being less smooth than the collinear one.  The second is the  opposite of (\textbf{P3}) which is that components with finite degree of smoothness in the two center solution are modified in a manner that reduces their degree of smoothness; again resulting in the horizon of the coplanar solution being less smooth than the collinear one.

In this paper, we take up the question of the degree of smoothness of horizons in $k$-center solutions, \emph{for all $k$}. Here $k$ is the number of centers in generic positions. Similar to the two and three center cases, it turns out one can allow for an arbitrary number of centers all distributed on a $\mathbf{R}^{k-1}$ subspace of the transverse Euclidean space. The $k$-center solution has a  spatial rotational symmetry $\mathbf{so}(d-k)$ in the black hole case and a $\mathbf{so}(9-k)$ in the membrane case.  To have a non-trivial spatial rotational isometry, we need that the number of centers in generic positions $k \leq d-2$ for the black hole case and $k \leq 7$ for the membrane case. When $k \geq d-1$ for the black hole case and $k \geq 8 $ in the membrane case, the solution has no spatial rotational isometries at all;  the only isometries are the the $\frac{\partial}{\partial t}$ static isometry in the black hole case and the $\frac{\partial}{\partial t}, \frac{\partial}{\partial x}, \frac{\partial}{\partial y}$ brane translation isometries in the membrane case.  These are the ``most generic multi center solutions'' that appear in the title; we will refer to this sometimes also as the ``$\infty$-center'' solution, $k = \infty$ is nothing but $k \geq d-1$ for the black hole case and $k \geq 8 $ in the membrane case.

The observations  described above, about how the horizon smoothness of the three center solution is related to the horizon of smoothness of the two center solution, can be used to draw lessons for the horizon smoothness of $k$-center solutions.  Before that, we will recall the key tool of organizing  in terms of what we call generalized Gegenbauer polynomials, first introduced in \cite{Gowdigere:2014aca}, which proves to be useful  in more ways than one.  We first introduce co-ordinates on the transverse Euclidean space, 
\ben 
x_1 &=& r  \cos \theta_1, \nonumber \\
x_2 &=& r \sin \theta_1 \, \cos \theta_2, \nonumber \\
\vdots \nonumber \\ 
x_{d-2} &=& r \sin \theta_1 \, \sin \theta_2 \, \sin \theta_3 \,\ldots \ldots \sin \theta_{d-3}\, \cos \theta_{d-2}, \nonumber \\ 
x_{d-1} &=& r \sin \theta_1 \, \sin \theta_2 \,\sin \theta_3 \,\ldots \ldots \sin \theta_{d-3}\, \sin \theta_{d-2},  \label{spcoordinates}
\een
in which the flat metric takes the form
\be\label{flatmetric}ds^2_{\mathbf{R}^{d-1}} = dr^2 + r^2 d\theta_1^2 + r^2\,\sin^2\theta_1\,d\theta_2^2 +\ldots  +  r^2\sin^2\theta_1\, \ldots\sin^2\theta_{d-3}\,d\theta_{d-2}^2  \, . \ee
Thus, the co-ordinates in the isotropic co-ordinate system are $t, r, \theta_1, \theta_2, \ldots \theta_{d-2}$. Note that the isotropic co-ordinate system is one in which the metric takes the form as in \eqref{rnsoln}. Different co-ordinate systems for the transverse Euclidean space, different from \eqref{spcoordinates}, \eqref{flatmetric} can also be used and we will need them later  (see \ref{214}, \ref{304})  for further discussion.

One then reorganises the harmonic function for the most generic solution \eqref{rnsoln2} as follows. First, choose the first black hole, the one with charge $\mu_1$ to be at the origin in the transverse Euclidean space and whose horizon we will study, i.e. choose $\vec{R}^{(1)} = 0$ in \eqref{rnsoln2}. The other black holes' centers  have generic co-ordinate positions: $\vec{R}^{(J)} \equiv( R_1^{(J)}, R_2^{(J)}, \ldots R_{d-1}^{(J)}), \quad J = 2, 3, \ldots$. Define for each black hole other than the first one, 
\ben \label{fR}
f^{(J)}(\theta_1, \ldots \theta_{d-2})  = \frac{R_1^{(J)}}{\|\vec{R}^{(J)} \|} \cos \theta_1 +\frac{R_2^{(J)}}{\|\vec{R}^{(J)} \|} \sin \theta_1 \cos \theta_2 + \ldots +\frac{R_{d-1}^{(J)}}{\|\vec{R}^{(J)} \|} \sin \theta_1  \ldots \sin \theta_{d-2} \een 
where
\ben
\|\vec{R}^{(J)} \| = +\sqrt{(R_1^{(J)})^2+(R_2^{(J)})^2+ \ldots +(R_{d-1}^{(J)})^2}
\een
is the Euclidean distance from the $J$'th black hole to the first one. $f^{(J)}(\theta_1, \ldots \theta_{d-2})$ is the cosine of the angle between the position vector $\vec{R}^{(J)}$of the $J$'th black hole and $\vec{r}$, the argument of the harmonic function.  The harmonic function \eqref{rnsoln2} for the most generic solution, can now be written as  
\be \label{3cenharm2}
H(r,\theta_1, \ldots \theta_{d-2}) = 1 + \frac{\mu_1}{r^{d-3}} + \sum_{J=2}^{\infty} \frac{\mu_J}{(\, r^2 - 2\, r\, \|\vec{R}^{(J)} \| f^{(J)}(\theta_1, \ldots \theta_{d-2}) + \|\vec{R}^{(J)} \|^2\,)^{\frac{d-3}{2}}}\,.
\ee
To further process the formula \eqref{3cenharm2}, we need the generating function of the $d$-dimensional Gegenbauer polynomials $G_n$\footnote{\label{fifth}We will not indicate the dimension in the notation of the Gegenbauer polynomials and also in the notation of the generalized Gegenbauer polynomials to avoid cluttering. The dimension can be read off from the context.} 
\be \label{gpgenerating} \frac{1}{(1- 2yz + z^2)^{\frac{d-3}{2}}} =  \sum_{n=0}^\infty  z^n \,G_n(y). \ee
Using \eqref{gpgenerating},  \eqref{3cenharm2} can be written as follows:
\be \label{3cenharm7}
H(r,\theta_1, \ldots \theta_{d-2}) = 1 + \frac{\mu_1}{r^{d-3}} + \sum_{J=2}^{\infty} \sum_{n=0}^{\infty} r^n\,\frac{\mu_J}{\|\vec{R}^{(J)} \|^{n+d-3}}\,G_n(f^{(J)}(\theta_1, \ldots \theta_{d-2})).
\ee
Now, we define generalized Gegenbauer polynomials\cref{fifth}
\be \label{ggpdefn}
{\cal G}_n (\theta_1, \ldots \theta_{d-2}) =  \delta_{n,0} + \sum_{J=2}^{\infty} \frac{\mu_J}{\|\vec{R}^{(J)} \|^{n+d-3}}\,G_n(f^{(J)}(\theta_1, \ldots \theta_{d-2})),
\ee
using which we can write the $r$-series expansion of the harmonic function \eqref{rnsoln2}, \eqref{3cenharm2} as follows:
\be \label{3cenharm3}
H(r,\theta_1, \ldots \theta_{d-2}) = \frac{\mu_1}{r^{d-3}} + \sum_{n = 0}^\infty r^n \, {\cal G}_n\,(\theta_1, \ldots \theta_{d-2})\,.
\ee
The term generalized Gegenbauer polynomials  was introduced in \cite{Gowdigere:2014aca} to  indicate such functions of the angular variables;  it is just a name and is not meant to indicate a new special function or anything else; in fact the main ingredient that goes into the construction of the generalized Gegenbauer polynomials is the Gegenbauer polynomial.  Note that a generalized Gegenbauer polynomial ${\cal G}_n (\theta_1, \ldots \theta_{d-2})$ contains in it's definition the charges and co-ordinate positions of all the black holes other than the first one whose horizon we are studying. It is thus a compact notation. The formula for the harmonic function in terms of the generalized Gegenbauer polynomials \eqref{3cenharm3} hides from view all these constants, making computations with this as the starting point, much cleaner.  What is more remarkable is that the results of the computations viz. the transition functions to the Gaussian null co-ordinate system and the components of the tensor fields in the Gaussian null co-ordinate system, are also expressed in terms of  these generalized Gegenbauer polynomials  albeit of the Gaussian null co-ordinates ${\cal G}_n (\Theta_1, \ldots \Theta_{d-2})$ and their partial derivatives. The compactness inherent in the notation of generalized Gegenbauer polynomials now translates into brevity for the final answers. Thus the use of these generalized Gegenbauer polynomials first of all makes the computations cleaner and easier and then allows us to express and report the results in a compact manner.  Note that the zeroth generalized Gegenbauer polynomial ${\cal G}_0$ is just a constant 
\be \label{ggp0} {\cal G}_0 (\theta_1, \ldots \theta_{d-2}) =  1 + \sum_{J=2}^{\infty} \frac{\mu_J}{\|\vec{R}^{(J)} \|^{d-3}}\,
\ee and the first generalized Gegenbauer polynomial ${\cal G}_1$ 
\begin{multline} \label{ggp1} {\cal G}_1 (\theta_1, \ldots \theta_{d-2}) = (d-3)\,\sum_{J=2}^{\infty} \frac{\mu_J}{\|\vec{R}^{(J)} \|^{d-1}}\,\left[R_1^{(J)} \, \cos \theta_1 +R_2^{(J)} \, \sin \theta_1 \cos \theta_2 + \ldots \right. \\ \left.+\,R_{d-1}^{(J)} \, \sin \theta_1 \ldots \sin \theta_{d-2} \right]
\end{multline}
is a non-constant function of the angles.  One can think of it as a linear combination of the $d-1$ functions $\cos \theta_1,~ \sin \theta_1 \cos\theta_2, \ldots$ $\sin \theta_1\,\sin \theta_2 \ldots \sin \theta_{d-2}$. We started with a certain co-ordinate system for the $S^{d-2}$ in the transverse Euclidean space, given in \eqref{spcoordinates} and we arrived at the above defined generalized Gegenbauer polynomials \eqref{fR} \eqref{ggpdefn}  and these particular summands in ${\cal G}_1$ \eqref{ggp1}. If one were to start with a different co-ordinate system for the transverse sphere, as we will in \ref{214} and \ref{304}, we would have analagous definitions of generalized Gegenbauer polynomials; ${\cal G}_1$ would still be a sum of $d-1$ summands but different to the ones appearing in \eqref{ggp1}.

For the two center case,  it is easiest\footnote{\label{perverse} In the two center case, one can think of aligning the two centers along an axis other than the $x_1$-axis. Then the generalized Gegenbauer polynomial would be a function of more than one angle and ${\cal G}_1$ would still  contain only one of the summands in \eqref{ggp1}. One can also think of aligning the two centers on a generic line away from any of the $x_i$-axes in which case the generalized Gegenbauer polynomial would be a function of all the angles and ${\cal G}_1$ would contain $d-1$ summands. Similarly for any $k$, one can align the black holes (i) in a way such that the generalized Gegenbauer polynomial is a function only of the first $k-1$ angles and ${\cal G}_1$ is a linear sum of the first $k-1$ summands in \eqref{ggp1} or (ii) in a way such that the generalized Gegenbauer polynomial is a function of more than $k-1$ angles and ${\cal G}_1$ is a linear sum of some $k-1$ summands in \eqref{ggp1} or (iii) in a generic way such that the generalized Gegenbauer polynomial is a function of all the angles and ${\cal G}_1$ is a lnear sum of all the $d-1$ summands in \eqref{ggp1}. Thus, the generalized Gegenbauer polynomials defined in \eqref{ggpdefn} for different values of the co-ordinate positions $R^{(J)}_l$ cover any and all $k$-center cases.} to line up the black holes on the $x_1$-axis. Then the harmonic function is a function of $r$ and $\theta_1$. Furthermore, the functions $f^{(J)}$ \eqref{fR} are functions of only one angle viz. $\theta_1$, the generalized Gegenbauer polynomial is nothing but a constant times the Gegenbauer polynomial of $\cos \theta_1$. In fact there is nothing much to gain by introducing generalized Gegenbauer polynomials and one can solve the problem otherwise \cite{Candlish1}. For the three center case, it is easiest \cref{perverse} to arrange the black holes on the $x_1 - x_2$ plane. Then the harmonic function is a function of $r$ and $\theta_1, \theta_2$.  The generalized Gegenbauer polynomials are functions of the two angles $\theta_1, \theta_2$ and ${\cal G}_1$ comprises only two summands viz. $\cos \theta_1$ and $\sin \theta_1 \cos \theta_2$. Here, the compactness afforded by the rewriting in terms of generalized Gegenbauer polynomials proves crucial to solve and report the results \cite{Gowdigere:2014aca}. Going on, for the $k$-center case ($k \leq d-2$), it is easiest\cref{perverse} to arrange them in the subspace spanned by $x_1$, $x_2\ldots x_{k-1}$ axes.  The harmonic function is a function of $r$ and the angles $\theta_1, \theta_2, \ldots \theta_{k-1}$; the generalized Gegenbauer polynomials are functions of the angles $\theta_1, \theta_2, \ldots \theta_{k-1}$ and ${\cal G}_1$ comprises the first $k-1$ summands in \eqref{ggp1}. We will sometimes refer to these as the $k$-center generalized Gegenbauer polynomials. Finally, for the most generic solution, the $\infty$-center case, the generalized Gegenbauer polynomials are functions of all the angles and ${\cal G}_1$ is given by \eqref{ggp1} comprising of all the $d-1$ summands.

The precise observations we made in \cite{Gowdigere:2014aca} about how the smoothness of the horizon in the three center solution is related to the one in the two center solution, which we have reviewed here (\textbf{(P1)}, \textbf{(P2)}, \textbf{(P3)}), leads one to assume that the smoothness of the horizon in the $k+1$-center solution is perhaps related to the one in the $k$-center solution in exactly the same way. Let us work out the consequences of this assumption for $k =3$, and for example $d\geq 6$ black holes.  We know the degree of smoothness of all the components of all tensor fields for the three center solution:  the metric components in footnote \ref{third} are clearly smooth, $g_{vv}$ is $\mathcal{C}^3$, $g_{v\Theta_i}$ for $i = 1,2$ are $\mathcal{C}^2$, all diagonal $g_{\Theta_i \Theta_i}$ and $g_{\Theta_1 \Theta_2}$ are $\mathcal{C}^1$, $A_{\lambda}$ is $\mathcal{C}^0$, $A_v$ is $\mathcal{C}^2$ and $A_{\Theta_i}$ for $i=1,2$ are $\mathcal{C}^0$ functions; all other components vanish and hence are $\mathcal{C}^\infty$. Thus the metric is $\mathcal{C}^1$ and gauge field $\mathcal{C}^0$. In the four center solution, the following additional components will be non-zero: $g_{v\Theta_3}$, $g_{\Theta_1 \Theta_3}$, $g_{\Theta_2 \Theta_3}$ and $A_{\Theta_3}$. If the above assumption we make is true, then it follows that the components  $g_{vv}$, $g_{v\Theta_i}$ for $i = 1,2$, all diagonal $g_{\Theta_i \Theta_i}$, $g_{\Theta_1 \Theta_2}$, $A_{\lambda}$, $A_v$ and $A_{\Theta_i}$ for $i=1,2$ all follow \textbf{(P3)}. The components given in footnote \ref{third} follow \textbf{(P1)}. Our assumption implies that the components $g_{v\Theta_3}$, $g_{\Theta_1 \Theta_3}$, $g_{\Theta_2 \Theta_3}$ and $A_{\Theta_3}$ will follow either \textbf{(P1)} or \textbf{(P2)}. But the tensor transformation law suggests it is \textbf{(P2)}. The assumption then implies that $g_{v\Theta_3}$, $g_{\Theta_1 \Theta_3}$, $g_{\Theta_2 \Theta_3}$ are at worst $\mathcal{C}^1$ functions while $A_{\Theta_3}$ is at worst $\mathcal{C}^0$.  Thus our assumption that the tensor components in the $k+1$-center solution are related to the ones in the $k$-center solution by only \textbf{(P1)}, \textbf{(P2)} or \textbf{(P3)},  provides us with an expectation for the series expansions and hence for the degrees of differentiabilities of all tensor components in the four center solution and consequently an expectation for the horizon smoothness. A similar exercise for $k=4$ provides an expectation for the five center solution and so on till we arrive at an expectation for the series expansions for all tensor components in the most generic solution, the $\infty$-center solution. In particular, we expect that the degree of smoothness of the horizon in the $\infty$-center solution is identical to that of the two center solution.

In the rest of this paper, we perform the computations to see if the above expectations are realized.  Clearly the problem is technically more complicated than the two and three center computations.  Due to the generic positioning of  the centers and the consequent absence of Killing symmetries in the transverse Euclidean space, there are virtually no first integrals available to make the job of solving the geodesic equations easier. One has to solve $d-1$ non-linear coupled differential equations for the $d-1$ functions $r(\lambda), \theta_1(\lambda), \theta_2(\lambda), \ldots, \theta_{d-2}(\lambda)$. The starting point of the computations in terms of generalized Gegenbauer polynomials makes the computations doable. Still, the task is quite formidable as it stands. But one realizes that to compute the degree of differentiability of any tensor component one only needs to compute a few low number of orders till one obtains the first fractional power of the affine parameter; these few low orders are controlled by only a few low orders in the series expansions of the transition functions.  Hence one would need to solve the geodesic equations only up to a certain point. Even before starting to solve the geodesic equations, we work out which coefficients in the series expansions are needed to check for all the expectations we have been provided.  For example, for $d=5$ black holes, it turns out that we only need to have the first six coefficients in the expansion of $r(\lambda)$ and only the first three non-trivial coefficients in the series expansion of each of $\theta_i(\lambda)$. It turns out this smaller task of solving the geodesic equations only up to the point required to determine the degree of horizon smoothness is quite simple, even doable by hand. After obtaining the transition functions, we compute the tensor components in the Gaussian null co-ordinate system and see if and how the expectations we have are realized. We try to isolate those aspects of the computations which are most relevant as answer to the question:  Do all $k$-center solutons have identical horizon smoothness and if so, why?

The rest of this paper is organized as follows. In section two \ref{2}, we study  the most generic multi center black holes first for $d = 5$ in \ref{21},  and then for all $d \geq 6$ in \ref{22}. We set up the computation of the horizon co-ordinate system in \ref{211} and \ref{221} and work out how many coefficients in the series expansions for the transition function  we would need  to check for our expectations.  We then solve the geodesic equations to the required order in \ref{212} and \ref{222}. Then, we compute the tensor components in the Gaussian null co-ordinate system in \ref{213} and \ref{223} and check for the expectations above. In \ref{214}, for only the $d=5$ case, we repeat all the computations with a different starting point viz. a different isotropic co-ordinate system and obtain results that corroborate the ones in \ref{213}. Then, in section three \ref{3}, we study the most generic multi center $M2$ brane horizons along the same lines as the black hole case and check for the above expectations in \ref{306}. In \ref{307} we work in an alternate isotropic co-ordinate system and obtain results identical to \ref{306}.  Finally, we conclude in \ref{4} with a  summary of the results.

\section{\label{2}The most generic multi center black holes}

The multi center black holes we investigate in this paper are solutions to $d$ dimensional Einstein-Maxwell theory, whose action is given by
\be \label{emaction}
S = \int d^dx~\sqrt{-g}\,\left(R - \frac{d-2}{8(d-3)}\,F_{\mu\nu}\,F^{\mu\nu} \right) \, .
\ee
We are following the conventions of  \cite{Candlish1} here. Following is the solution in  isotropic co-ordinates: the metric and gauge fields are given by
\be \label{rnsoln}
ds^2 = -  H^{-2}\,dt^2 + H^{\frac{2}{d-3}}\, ds^2_{\mathbf{R}^{d-1}}, \qquad  A = -\frac{dt}{H}\, , \ee
where $ds^2_{\mathbf{R}^{d-1}}$ is the flat metric of the transverse Euclidean space $\mathbf{R}^{d-1}$. $H$ is a harmonic function in the transverse Euclidean space:
\be  \label{rnsoln2}H(\vec{r}) = 1 + \sum_{J = 1}^\infty \frac{\mu_J}{\| \vec{r} - \vec{R}^{(J)} \|^{d-3}} \, .\ee 
$\vec{R}^{(i)}$ are points in the transverse Euclidean space which  correspond to the locations of the horizons of the various black holes and $\| \|$ is the Euclidean norm. 

In the following, we will implement the procedure to study horizon smoothness, already described in the introduction, first for five dimensional black holes which behave differently to the six and higher dimensional black holes whose study we take up  subsequently. 

\subsection{\label{21}$d = 5$}
We start by setting $d= 5$ in all previous formulae \eqref{spcoordinates}-\eqref{ggp1}; in particular, the harmonic function is\cref{fifth}
\be \label{5dharm}
H(r,\theta_1, \theta_2, \theta_3) = \frac{\mu_1}{r^{2}} + \sum_{n = 0}^\infty r^n \, {\cal G}_n\,(\theta_1, \theta_2, \theta_3)\,.
\ee
\subsubsection{\label{211}Constructing the Gaussian null co-ordinate system}

As already described in the introduction, the horizon co-ordinate system of choice is the Gaussian null co-ordinate system, whose constuction needs the solution to the geodesic equations. 

The only Killing symmetry of the metric is $\frac{\partial}{\partial t}$, due to which the ``$t$-geodesic'' equation admits a first integral which can be solved, 
\ben \label{5dtgeode}\frac{d}{d\lambda} \left[ H^{-2} \, \frac{dt}{d\lambda} \right] = 0 ~&\Longrightarrow& ~ \frac{d}{d\lambda}t(\lambda) = -  H(r(\lambda), \theta_1(\lambda), \theta_2(\lambda),\theta_3(\lambda))^2 \nonumber \\ ~ &\Longrightarrow& ~  t(\lambda) = v \,- \int d\lambda \,H(r(\lambda), \theta_1(\lambda), \theta_2(\lambda),\theta_3(\lambda))^2, \een
where in choosing the integration constant of the first integration  to be $-1$ we have  employed some of the freedom in choosing the affine parameter and $v$ is the second integration constant. Thus, $t(\lambda)$ is determined via \eqref{5dtgeode} in terms of $r(\lambda), \theta_1(\lambda), \theta_2(\lambda)$ and $\theta_3(\lambda)$,  which are obtained by solving simultaneously the other geodesic  equations. We will solve the ``$\theta_i$-geodesic'' equations, for $i = 1, 2, 3$:
\be \label{5dthetaigeod}\ddot{\theta_i} - \frac{\partial_{\theta_i} H}{r^2 \, F_i(\theta_1,\theta_2, \theta_3)} - \frac{\partial_{\theta_i} H}{2 H r^2 \, F_i(\theta_1,\theta_2, \theta_3)}\,\dot{r}^2  + \frac{\partial_r H }{H}\,\dot{r}\,\dot{\theta_i} + \frac{2}{r}\,\dot{r}\,\dot{\theta_i}  + \ldots = 0,\ee
where 
\be \label{5dFi} F_1(\theta_1, \theta_2, \theta_3) = 1, \quad F_2(\theta_1, \theta_2, \theta_3) =  \sin^2 \theta_1, \quad F_3(\theta_1, \theta_2, \theta_3) =  \sin^2 \theta_1 \, \sin^2 \theta_2\ee
and the null condition:
\be \label{5dnull}- H^{-2}\,\dot{t}^2 + H\,\dot{r}^2 + H r^2\,\dot{\theta_1}^2 + Hr^2\,\sin^2\theta_1\, \dot{\theta_2}^2 + Hr^2\,\sin^2\theta_1\,\sin^2\theta_2\,\dot{\theta_3}^2 = 0,\ee
after using \eqref{5dtgeode} becomes
\be \label{5dnull1}- H + \dot{r}^2 +  r^2\,\dot{\theta_1}^2 + r^2\,\sin^2\theta_1\,
\dot{\theta_2}^2 + r^2\,\sin^2\theta_1\,\sin^2\theta_2\,\dot{\theta_3}^2 = 0.\ee
The boundary conditions are chosen as follows. First we employ the remaining freedom allowed in choosing the affine parameter so that the affine parameter takes the value zero at the horizon of the first black hole and the part of the geodesic that lies outside this horizon in the isotropic co-ordinate patch corresponds to $\lambda > 0$.  Since the isotropic co-ordinate $r$ is such that it limits to the value zero as one approaches the horizon of the first black hole, we should impose the following boundary condition for $r(\lambda)$: 
\be \label{5drbound} r(\lambda = 0) = 0.\ee
The geodesics in question are  purely radial geodesics without any angular momentum; hence the boundary conditions for the angles are 
\be \label{anglebound} \theta_i(\lambda = 0) = \Theta_i, \qquad \dot{\theta_i}(\lambda = 0) = 0,\qquad  i  = 1, 2, 3   \ee
where $\Theta_i$ are arbitrary constants at this stage.

The equations \eqref{5dthetaigeod},\eqref{5dnull1} are highly non-linear coupled equations and are probably impossible to solve directly. The strategy adopted \cite{Candlish1}  is to assume a series expansion for each of the unknown functions $r(\lambda)$, $\theta_i(\lambda)$. The expansion parameter is an appropriate power of the affine parameter $\lambda$ and it can be motivated as follows.  We  compute the behavior of $r(\lambda)$  near the horizon by examining the leading (in $\lambda$) behavior of  the null condition, which is:
\be \dot{r}^2 =  H \quad \Longrightarrow \quad \dot{r}^2 \sim \frac{1}{r^{2}} \quad \Longrightarrow \quad  r(\lambda)^2 \sim \lambda \qquad \Longrightarrow \qquad r(\lambda) \sim \sqrt{\lambda}.\ee
This together with a similar examination of the behavior of the $\theta_i$-geodesic equations near the horizon, motivates the following series expansion ansatz: 
\be r(\lambda) = \sum_{n = 0}^\infty c_n \, \left(\sqrt{\lambda} \right)^n,\qquad \theta_i(\lambda) = \sum_{n = 0}^\infty b^{(i)}_n \, \left(\sqrt{\lambda} \right)^n, \qquad i  = 1, 2, 3. \ee
The boundary conditions \eqref{5drbound} and \eqref{anglebound}  then imply that the following co-efficients vanish: \be c_0 = 0, \quad b^{(i)}_1 =  0,\quad b^{(i)}_2 = 0.  \ee 
We thus have 
\be \label{expansionansatz} r(\lambda) = \sum_{n = 1}^\infty c_n \, \left(\sqrt{\lambda} \right)^n,\quad \quad \theta_i(\lambda) = \Theta_i + \sum_{n = 3}^\infty b^{(i)}_n \, \left(\sqrt{\lambda} \right)^n.\ee
The procedure to obtain the solutions to the geodesic equations \cite{Candlish1} is to plug in the expansions \eqref{expansionansatz} into the geodesic equations, obtain a series expansion of the equations in $\sqrt{\lambda}$ and solve order by order. One would obtain the coefficients $c_n$'s and the $b^{(i)}_n$'s as functions of the constants $\Theta_i$. The solutions to the geodesic equations are hence  functions of the affine parameter $\lambda$ and the constants: $r( \lambda, \Theta_1, \Theta_2, \Theta_3), \theta_i( \lambda, \Theta_1, \Theta_2, \Theta_3)$. One then uses \eqref{5dtgeode} to obtain 
\ben \label{228} t(\lambda)  &=& v \,- \int d\lambda \,H(r(\lambda), \theta_1(\lambda), \theta_2(\lambda),\theta_3(\lambda))^2 \nonumber \\ &\equiv& v \,- T(\lambda, \Theta_1, \Theta_2, \Theta_3) \een
These solutions to the geodesic equations, which are functions of the affine parameter $\lambda$ and the constants $v, \Theta_1, \Theta_2, \Theta_3$, are the transition functions between the isotropic co-ordinates $t, r, \theta_1, \theta_2, \theta_3$ and the Gaussian null co-ordinates $\lambda, v, \Theta_1, \Theta_2, \Theta_3$.

Before we implement this procedure we will ask ourselves the question: What is the minimal number of the $c_n$'s and the $b^{(i)}_n$'s  needed to check for the expectations one has for  the horizon smoothness of the most generic solution? 

$g_{vv}$ is $\mathcal{C}^3$ in the three center solution i.e. the first fractional power in its series expansion is $\lambda^{7/2}$ at order seven. Hence when going form three to four and subsequently in every step one expects it to follow \textbf{(P3)} which means that in the in the most generic solution it is expected to have a series expansion with first fractional power $\lambda^{7/2}$. To be able to compute to this order, from\eqref{5dgvvbefore}, we need only  the coefficients $c_1 - c_4$. $g_{v\Theta_1}$ and $g_{v\Theta_2}$ are $\mathcal{C}^2$ in the three center solution i.e. the first fractional power in its series expansion is $\lambda^{5/2}$ at order five. Hence when going form three to four and subsequently in every step one expects it to follow \textbf{(P3)} which means that in the in the most generic solution it is expected to have a series expansion with first fractional power $\lambda^{5/2}$. To be able to compute to this order, from\eqref{5dgvtbefore}, we need only  the coefficients $c_1 - c_4$. $g_{v\Theta_3}$ is vanishing and hence $\mathcal{C}^\infty$ in the three center solution. It is expected to follow \textbf{(P2)} which means that in the most generic solution it is expected to have a series expansion with first fractional power $\lambda^{5/2}$. To be able to compute to this order, from\eqref{5dgvtbefore}, we need only  the coefficients $c_1 - c_4$. $g_{\Theta_i \Theta_i}$ for all $1 \leq i \leq 3$ are $\mathcal{C}^2$ in the three center solution  i.e. the first fractional power in its series expansion is $\lambda^{5/2}$ at order five. Hence they are expected to follow \textbf{(P3)} which means that in the in the most generic solution it is expected to have a series expansion with first fractional power $\lambda^{5/2}$. To be able to compute to this order, from \eqref{5dgtt11befor}  \eqref{5dgtt22before} and \eqref{5dgtt33before}, we need only  the coefficients $c_1 - c_6$ and $b^{(i)}_3 - b^{(i)}_5$. $g_{\Theta_1 \Theta_2}$ in the three center solution is $\mathcal{C}^2$. Hence it is expected to follow \textbf{(P3)}.  $g_{\Theta_1 \Theta_3}$ and $g_{\Theta_2 \Theta_3}$ are vanishing in the three center and hence expected to follow \textbf{(P2)}. Hence to compute to this order for $g_{\Theta_i \Theta_j}$ for each $1 \leq i\neq j \leq 3$,   from \eqref{5dgtt12before}, \eqref{5dgtt13before} and \eqref{5dgtt23befor},  we only need the co-efficients  $c_1 - c_6$, $b^{(i)}_3 - b^{(i)}_5$ and $b^{(j)}_3 - b^{(j)}_5$.  To check the expectations for the components of the gauge field, it follows  from \eqref{5dAbefore}, it follows that we need no more than the coefficients $c_1 - c_4$.

To conclude, we set ourselves the much reduced goal of solving the geodesic equations only upto the point needed to obtain $c_1 - c_6$ and $b^{(i)}_3 - b^{(i)}_5$ for each $i=1,2,3$.

\subsubsection{\label{212}Solving the geodesic equations}

We now solve the geodesic equations. It is convenient to solve the $\theta_i$-geodesic equations \eqref{5dthetaigeod} together with the null condition \eqref{5dnull1}. We will see that there is a decoupling of sorts that happens: the coefficients $c_1-c_6$ are determined by the null condition, the coefficients $b^{(i)}_3-b^{(i)}_5$ are determined by the $\theta_i$-geodesic equation.

\textbf{Null condition:} We start with the analysis of the null condition \eqref{5dnull1}. Using  \eqref{expansionansatz}, we can work out the $\sqrt{\lambda}$-series expansion of the (left hand side of the) null condition. The last three terms, viz. $ H r^2\,\dot{\theta_1}^2 + Hr^2\,\sin^2\theta_1\, \dot{\theta_2}^2 + Hr^2\,\sin^2\theta_1\,\sin^2\theta_2\,\dot{\theta_3}^2$ start at order four while the first two terms start at order minus two. Hence the first six non-trivial orders of the null condition, which are the orders from minus two to plus three, receive contributions from only the first two terms. The coefficeints in the $\sqrt{\lambda}$-expansion of the second term i.e. $\dot{r}^2$ are clearly  functions of the $c_n$'s only;  it is easy to see that the first six non-trivial orders are functions of the terms $c_1 - c_6$. Hence the contribution of  $\dot{r}^2$ to the orders from minus two to plus three contain precisely the $c_n$ coefficients we need to solve for. Similarly, we will see that the contribution of the first term i.e. $-H$ to orders from minus two to plus three also contain only those $c_n$ coefficients that we need to solve for and no other coefficient. First, from \eqref{5dharm}, we can see that contributions from orders minus two to plus three come from the first black hole term $\frac{\mu_1}{r^2}$ and only from the first four terms in the summation i.e. ${\cal G}_0, r{\cal G}_1, r^2{\cal G}_2$ and $r^3{\cal G}_3$. The first black hole term's contribution to orders minus two to plus three will contain functions of the $c_n$'s only; in fact they will be functions of the required $c_1 - c_6$. Clearly ${\cal G}_0$ is a constant and contributes only to order zero. Now consider $r {\cal G}_1$. It's $\sqrt{\lambda}$-series expansion starts off from order one and since ${\cal G}_1$ is a function of the isotropic angles, the coefficients could involve the $b^{(i)}_n$'s also. But a closer examination (using \eqref{expansionansatz}) reveals that $b^{(i)}_n$'s start appearing only from order four onwards. Similarly in the $\sqrt{\lambda}$-series expansion of $r^2 {\cal G}_2$ and $r^3 {\cal G}_3$, the $b^{(i)}_n$'s start appearing only from order five and order six onwards respectively. Thus, we can see that the contribution of $-H$ to orders minus two to plus three are functions of only the required $c_1 - c_6$, with none of the $b^{(i)}_n$'s making an appearance.

This means that we only need to examine the first six non-trivial orders of the null condition from orders minus two to plus three to obtain the required coefficients $c_1 - c_6$. It turns out that at order minus two only $c_1$ occurs and hence gets determined. Then at order minus one $c_2$ occurs linearly and gets uniquely determined. At every successive order, the successive coefficient  occurs linearly and gets uniquely determned. We can do all this readily by hand (no need of any computer algebra package) and obtain:
\begin{multline} \label{5drtfo} r(\lambda, v, \Theta_1, \Theta_2, \Theta_3)  = \sqrt{2}\mu_1^{1/4}\,\lambda^{1/2} + \frac{1}{2\sqrt{2}\mu_1^{1/4}}{\cal G}_0 \,\lambda^{3/2} 	+ \frac{2}{5}{\cal G}_1\,\lambda^2  -\frac{1}{48 \sqrt{2} \mu _1^{3/4}}\left[ 3\, {\cal G}_0^2 - 32\mu_1\, {\cal G}_2  \right]  \lambda^{5/2} 	\\  -\frac{2}{35 \mu _1^{1/2}} \left[ {\cal G}_0 \,{\cal G}_1  -10 \mu_1\,{\cal G}_3 \right] \lambda^3 + \ldots   \end{multline}
In the above, ${\cal G}_n$'s appearing are all functions of the Gaussian null co-ordinate angles $\Theta_1, \Theta_2, \Theta_3$.

For any $k$-center solution, the result for $c_1 - c_6$ will still be given by \eqref{5drtfo}, with the understanding that one has to replace with generalized Gegenbauer polynomials appropriate for $k$-center solution, i.e. the ones with $k-1$ summands in \eqref{fR}. Hence the result \eqref{5drtfo} for \emph{ $c_1 - c_6$ is independent of $k$ (the number of arbitrarily positioned centers).} This feature has it's origin in the fact that up to this order in the computation none of the $b^{(i)}_n$'s show up. The $b^{(i)}_n$'s are accompanied by derivatives of the generalized Gegenbauer polynomials which will be different for different $k$.  This independence from $k$ of the results of $c_1 - c_6$ will feature in subsequent analysis. 

\textbf{$\theta_i$-geodesic equations :} We begin by working out the $\sqrt{\lambda}$-series expansion of the $\theta_i$-geodesic equations. The terms that we have not displayed in \eqref{5dthetaigeod} are the ones proportional to $\dot{\theta_j}\,\dot{\theta_k}$ and start from order two. It turns out that for the purpose of determining $b^{(i)}_3 - b^{(i)}_5$, it is enough to consider only up to order one. The terms displayed are the ones that contribute to the first three orders from order minus one to plus one. Evaluating these orders using \eqref{expansionansatz} shows that they are functions only of (i) the already determined coefficients $c_1-c_3$ and of (ii) $b^{(i)}_3 - b^{(i)}_5$, with none of the $b^{(j)}_n$'s for $j \neq i$ making an appearance (however, they do make an appearance from order two onwards). This is the decoupling alluded to earlier: for a given $i$, the required coefficients $b^{(i)}_3 - b^{(i)}_5$ appear (earliest in the series expansion) only in the $\theta_i$-geodesic equation for that $i$. Hence, it does not matter what order we solve the $\theta_i$-geodesic equations in, as long as we consider them after obtaining the solution to the null condition. We thus obtain
\begin{multline} \label{5dthetaitfo}
\theta_i(\lambda, v, \Theta_1, \Theta_2, \Theta_3) = \Theta_i + \frac{\sqrt{2} }{\mu _1^{1/4}} \frac{\partial_{\Theta_i} {\cal G}_1}{F_i(\Theta_1, \Theta_2, \Theta_3)}\,\lambda^{3/2} + \frac34 \frac{\partial_{\Theta_i} {\cal G}_2}{F_i(\Theta_1, \Theta_2, \Theta_3)} \,\lambda^2 \\   -\frac{1}{10 \sqrt{2} \mu _1^{3/4}} \frac{17 \,{\cal G}_0\, \partial_{\Theta_i} {\cal G}_1 - 8 \mu _1\, \partial_{\Theta_i} {\cal G}_3 }{F_i(\Theta_1, \Theta_2, \Theta_3)} \lambda^{5/2} + \ldots
\end{multline} 
where the $F_i(\Theta_1, \Theta_2, \Theta_3)$ are defined in \eqref{5dFi}.
Using the above, one can compute \eqref{228} and obtain
\be \label{5dttfo} t( \lambda, v, \Theta_1, \Theta_2, \Theta_3) = v - \, T(\lambda, \Theta_1, \Theta_2, \Theta_3) ,\ee
where 
\begin{multline} \label{5dTo}
T(\lambda, \Theta_1, \Theta_2, \Theta_3) = -\frac{\mu _1}{4}\,\lambda^{-1} + \frac{3\mu_1^{1/2}}{4}  {\cal G}_0 \log \lambda + \frac{8\sqrt{2} \mu _1^{3/4}}{5}  {\cal G}_1  \lambda^{1/2} + \frac{1}{48} \left[33 {\cal G}_0^2 + 80 \mu_1\,{\cal G}_2 \right] \lambda  \\+ \frac{2\sqrt{2}\mu_1^{1/4}}{35}\left[19 {\cal G}_1 {\cal G}_0 + 20\mu _1\, {\cal G}_3  \right] \lambda^{3/2} + \ldots 
\end{multline}
We have now obtained in \eqref{5drtfo}, \eqref{5dthetaitfo} and in \eqref{5dttfo} the minimally needed definition of the horizon co-ordinate system for the horion of the first black hole, with which we can check for the expecations we have for the degree of horizon smoothness.

\subsubsection{\label{213}Tensor components in Gaussian null co-ordinates}

Now that we have obtained the transition functions to the required order in \eqref{5drtfo} -\eqref{5dttfo}, we only need to substitute them in the tensor transformation law to obtain the tensor components in the Gaussian null coordinate system. For each component, we will compare the answer with the expectation for its degree of differentiability after having reviewed the expectation. First, let us dispense with those components which we do not have to evaluate.  From the definition of the Gaussian null co-ordinate system, it follows (see \cite{Gowdigere:2014aca}) that the following metric components are constant and hence smooth functions; we will not evaluate them. 
\be \label{lambdarow} g_{\lambda \lambda}= 0, ~g_{\lambda v} = 1, ~g_{\lambda \Theta_i} = 0 \ee
We will compute the following fifteen components, which are expected to be non-zero and not smooth: 
$g_{vv}$, $g_{v \Theta_i}$, $g_{\Theta_i \Theta_j}$, $A_{\lambda}$, $A_v$, $A_{\Theta_i}$, with $1 \leq i,j \leq 3.$ Note that in all the formulae appearing here in \ref{213} (and in \ref{212}), the generalized Gegenbauer polynomials are functions of the Gaussian null co-ordinate angles $\Theta_i$'s. For example,
\begin{multline} \label{ggp1gnco} {\cal G}_1  = \sum_{i=2}^{\infty} \frac{2 \mu_i}{\|\vec{R}^{(i)} \|^{4}}\,\left[R_1^{(i)} \, \cos \Theta_1 +R_2^{(i)} \, \sin \Theta_1 \cos \Theta_2 +R_3^{(i)} \, \sin \Theta_1 \sin \Theta_2 \cos \Theta_3 \right. \\ \left.+\,R_{4}^{(i)} \, \sin \Theta_1 \sin \Theta_2 \, \sin \Theta_{3} \right].
\end{multline}
\underline{First} consider the class of components: $g_{vv}$, $A_{\lambda}$, $A_v.$  Computations up to the required order of this class of components requires \eqref{5dgvvbefore},\eqref{5dAbefore} the expressions for $c_1 - c_4$ only (not even their derivatives). We obtain:
\ben \label{5dgvv} g_{vv} &=& -\frac{4}{\mu _1}\lambda^2 + \frac{12}{\mu _1^{3/2}}\, {\cal G}_0\,\lambda^3 + \frac{64 \sqrt{2}}{5 \mu _1^{5/4}}\,{\cal G}_1\,\lambda^{7/2} + \ldots \\ \label{5dAl} A_\lambda &=& \frac{\mu _1^{1/2}}{2}\lambda^{-1} + \frac{3}{4}{\cal G}_0 + \frac{4\sqrt{2} \mu _1^{1/4}}{5}  {\cal G}_1\lambda^{1/2} + \ldots \\ \label{5dAv} A_v &=& -\frac{2}{\mu _1^{1/2}} \lambda  + \frac{3 }{\mu _1} {\cal G}_0\, \lambda^2 + \frac{16 \sqrt{2} }{5 \mu _1^{3/4}} {\cal G}_1 \lambda^{5/2} + \ldots  \een
Recall that the expressions for $c_1-c_6$ are independent of $k$ in a certain way \ref{212};  that is, one can start with the answer for say two center or three center case (expressed in terms of generalized Gegenbauer polynomials) and to obtain the $k$-center answer one only has to replace with the generalized Gegenbauer polynomials relevant for the $k$-center solution (i.e. the one with $k-1$ summands in ${\cal G}_1$). This feature of the coefficients $c_1-c_4$ translates to the expressions for the tensor components $g_{vv}$, $A_{\lambda}$,  and $A_v$ as well;  we could just borrow the expressions for them from \cite{Gowdigere:2014aca} and be assured of having obtained the correct answer.  Note that these components are non-zero and non-smooth even in the two and three center cases.  Hence, according to our expectation, when going from three center to four center and in every subsequent step from $k$-center to $k+1$-center they follow \textbf{(P3)}, that is, they will be modified for sure, but there will not be any modification in the series expansion and hence no modification in the degree of differentiability. That is, for every $k$, these components have the same degree of differentiability.  Here, in the results \eqref{5dgvv}-\eqref{5dAv}, we see this expectation playing out. We can even state precisely the modification: it is simply the replacement of the generalized Gegenbauer polynomial relevant to the $k$-center solution with the the generalized Gegenbauer polynomial relevant to the $k+1$-center solution.

We then consider the \underline{second} class of components,  $g_{v \Theta_i}$, $A_{\Theta_i}$. Computations up to the required order of this class of components requires \eqref{5dgvtbefore},\eqref{5dAbefore} the expressions for $c_1 - c_4$ and their derivatives. We obtain:
\ben  \label{5dgvThetai} g_{v\Theta_i}  &=& -\frac{32 \sqrt{2}}{5 \mu _1^{1/4}}\,\partial_{\Theta_i} \,{\cal G}_1\, \lambda^{5/2}  + \ldots \\ \label{5dAThetai} A_{\Theta_i} &=& \frac{16 \sqrt{2} \mu _1^{1/4}}{5}  \partial_{\Theta_i} {\cal G}_1\, \lambda^{3/2}  + \ldots \een
For $i = 1, 2,$ these components are non-zero and non-smooth in the three center solution. According to our expectation, when going from three center to four center and in every subsequent step from $k$-center to $k+1$-center they follow \textbf{(P3)}. That is, they are modified without any modificication in  the degree of differentiability. We see these expectations playing out here. The modification is simply the replacement with the relevant generalized Gegenbauer polynomial. The absence of change in degree of differentiability is due to the fact that for all $k$-center solutions with $k \geq 3$, ${\cal G}_1$ is a function of both $\Theta_1$ and $\Theta_2$. For $i > 2$, the components $g_{v \Theta_i}$, $A_{\Theta_i}$ are zero in the three center solution. According to our expectation, when going step by step from three to four center to $\ldots$ etc, these components first follow either \textbf{(P1)} or \textbf{(P2)}. Subsequently after the first time \textbf{(P2)} is realized, they become non-zero and non-smooth, after which they follow \textbf{(P3)}. This means that these components are smooth till a certain stage (for some $k$-center solution) after which they become non-smooth without changing the overall differentiability of the tensor field. We see this expectation playing out in the results above. For a given $i$, the components \eqref{5dgvThetai} and \eqref{5dAThetai} are zero for all $k$-center solutions with $k \leq i$ and for $k > i$ they become non-smooth without changing the differentiability of the tensor field. 

Now we consider the \underline{third} and final class of components,  $g_{\Theta_i \Theta_j}$ with $1 \leq i, j \leq 3.$ Computations up to the required order of this class of components requires \eqref{5dgtt11befor}-\eqref{5dgtt23befor} the expressions for $c_1 - c_6$ and their derivatives, and for $b^{(i)}_3-b^{(i)}_5$ and their derivatives. We obtain for the diagonal components,
\begin{multline} \label{5dgii} \frac{g_{\Theta_i\Theta_i}}{F_i(\Theta_1, \Theta_2, \Theta_3)} =  \mu_1 +  2\mu_1^{1/2}   \, {\cal G}_0 \, \lambda + 2 \sqrt{2} \mu _1^{3/4} \,\Delta^{(ii)}( {\cal G}_1)\, \lambda^{3/2} +   \left[{\cal G}_0^2 + \frac{3 \mu_1}{2} \left(\Delta^{(ii)}+\frac{5}{3}\right)({\cal G}_2)  \right] \lambda^2\\  +\frac{4 \sqrt{2} \mu _1^{5/4}}{5}  \left[ \left(\Delta^{(ii)}+4\right)({\cal G}_3) \right] \lambda^{5/2}+ \ldots \end{multline}
and for the non-diagonal  components,
\begin{multline} \label{5dgij} g_{\Theta_i\Theta_j} =  2 \sqrt{2} \mu _1^{3/4} \,\Delta^{(ij)}( {\cal G}_1)\, \lambda^{3/2} +  \frac{3 \mu_1}{2} \Delta^{(ij)}({\cal G}_2)\, \lambda^2 +\frac{4 \sqrt{2} \mu _1^{5/4}}{5}  \Delta^{(ij)}({\cal G}_3)  \lambda^{5/2}+ \ldots \end{multline}
where the $\Delta^{(ij)}$ are the following six second order differential operators:
\be\Delta^{(ii)} = 1 + \frac{1}{F_i(\Theta_1, \Theta_2, \Theta_3)} \frac{\partial^2~~}{\partial \Theta_i^2} + \sum_{k=1}^{i-1} \frac{\cot \Theta_k}{F_k(\Theta_1, \Theta_2, \Theta_3)} \, \frac{\partial~~ }{\partial \Theta_k} \nonumber \ee
\be  \label{5dDeltao} \Delta^{(ij)} = \frac{\partial^2 ~~}{\partial \Theta_i  \partial \Theta_j} - \cot \Theta_i \frac{\partial ~~}{\partial \Theta_j}, \qquad i < j\ee
with the $F_i(\Theta_1, \Theta_2, \Theta_3)$ given in \eqref{5dFi}. The above $\Delta^{(ij)}$ are second order differential operators in  Gaussian null co-ordinates. The remarkable fact is that each of the functions of angles  that appear as summands in \eqref{ggp1gnco} is in the kernel of each of the $\Delta^{(ij)}$'s. Since ${\cal G}_1$  for different $k$-center solutions is a (different) linear combination of these functions of angles, ${\cal G}_1$  is in the kernel of each of the  $\Delta^{(ij)}$'s for all $k$-center solutions. Thus we have 
\be \label{kernel} \Delta^{(ij)}({\cal G}_1) = 0, \qquad \text{for all $k$-center solutions.}\ee
Note that the diagonal components of $g_{\Theta_i \Theta_i}$ and the off-diagonal component $g_{\Theta_1 \Theta_2}$ are non-zero in the three center solution. According to our expectation, when going from three center to four center and in every subsequent step from $k$-center to $k+1$-center they follow \textbf{(P3)}. That is, they are modified without any modificication in  the degree of differentiability. We see this expectation being played out in \eqref{5dgii}, \eqref{5dgij} because of \eqref{kernel}. The off-diagonal components $g_{\Theta_1 \Theta_3}$ and $g_{\Theta_2 \Theta_3}$ vanish and are smooth for the three center solution. Hence the expectation is that they follow \textbf{(P2)}, which means that they are ${\mathcal C}^2$ functions. This expectation is realized in \eqref{5dgij} again due to \eqref{kernel}.

With the aid of the actual computations of tensor components in the Gaussian null co-ordinate system, we are able to see that the surmise we made for the $k+1$-center horizon smoothness to the $k$-center horizon smoothness and the consequent expectations are all realized in reality. We have thus shown  that \emph{the horizon smoothness is identical for all $k$-center solutions including the most generic solution, the $\infty$-center solution}.

Not only is the horizon smoothness identical for all $k$-center solutions, we have seen that even an  individual component (whenever it has a finite degree of differentiability) has identical series expansions and hence identical degree of differentiability for all $k$-center solutions (for which it has a finite degree of differentiability). Let us try to gather why this happens, component by component.  
$g_{vv}$ is a $\mathcal {C}^3$ function for all $k$-center solutions, because all the odd orders in its $\sqrt{\lambda}$-expansion upto order five vanish \eqref{5dgvv}. And this vanishing is due to two reasons \eqref{5dgvvbefore}. The first is the series ansatz \eqref{expansionansatz} which is due to the boundary conditions \eqref{5drbound}, \eqref{anglebound}, which clearly are independent of $k$. The second reason is that in the solution to the geodesic equations \eqref{5drtfo},  $c_2 = 0$.  There is an independence of $k$ to the fact that $c_2 = 0$ in the solution to the geodesic equations. Recall from \ref{212} that for the coefficients $c_1-c_6$ the expressions are functions of the generalized Gegenbauer polynomials and it is the same expression for all $k$ albeit with the understanding that it is the generalized Gegenbauer polynomial relevant for that $k$. Now $c_2$ (and also $c_1$) is a constant and takes the same value for all $k$. Now  $g_{v\Theta_i}$ are $\mathcal{C}^2$ functions (whenever their degree of differentiability is finite) for all $k$-center solutions, because all the odd orders in its $\sqrt{\lambda}$-expansion upto order three vanish \eqref{5dgvThetai}. This vanishing is due to two reasons  \eqref{5dgvtbefore}, both independent of $k$; first again being the series ansatz \eqref{expansionansatz} and the second being  the fact that $c_1$ and $c_2$ are constant functions in the solution to the geodesic equations \eqref{5drtfo}, as opposed to the apriori possibility that they are non-trivial functions of the Gaussian null co-ordinate angles. The components $g_{\Theta_i \Theta_j}$ are $\mathcal{C}^2$ functions (whenever their degree of differentiability is finite) for all $k$-center solutions, because all the odd orders in the $\sqrt{\lambda}$-expansion up to order three vanish \eqref{5dgii}, \eqref{5dgij}. This vanishing is due to four reasons \eqref{5dgtt11befor}-\eqref{5dgtt23befor}, all independent of $k$. The first is again the series ansatz \eqref{expansionansatz}.  The second  is  the previously appeared fact that $c_1$ and $c_2$ are constant functions in the solution to the geodesic equations. The third is the fact that $c_3$, even if a different constant for different $k$, is a constant function, i.e. all its partial derivatives vanish. The second and third reasons cause the odd order terms up to order three in $- \frac{\partial_{\Theta_i}T\,\partial_{\Theta_j}T}{H^2} + \partial_{\Theta_i}r\,\partial_{\Theta_j}r\, H$ to vanish. The fourth reason is the appearance of the differential operators \eqref{5dDeltao} and the fact \eqref{kernel} which as we have noted is independent of $k$.  The component $A_{\lambda}$ is a $\mathcal{C}^0$ function \eqref{5dAl} for all $k$-center solutions. This happens again because of the boundary conditions and the independence from $k$ of $c_1$ and $c_2$. The component $A_v$ is a $\mathcal{C}^2$ function \eqref{5dAv} for all $k$-center solutions. This happens again because of the boundary conditions and the independence from $k$ of  the fact that $c_2=0$. The component $A_{\Theta_i}$ is a $\mathcal{C}^1$ function \eqref{5dAThetai} for all $k$-center solutions. This happens again because of the boundary conditions and the independence from $k$ of  $c_1$, $c_2$ and $c_3$.

In summary, the underlying reasons behind the statements: 
``The horizon smoothness identical for all $k$-center solutions, including the most generic solution, the $\infty$-center solution;  The degree of differentiability of individual tensor components (when it is finite) is identical for all $k$-center solutions''
seems to be the following three:
\begin{itemize}
\item \textbf{(R1)} The boundary conditions that determine the series ansatze for $r(\lambda)$, $\theta_i(\lambda)$ are identical for all $k$-center solutions. 
\item \textbf{(R2)} In the solution to the geodesic equations, the first three  coefficients in the series expansion for $r(\lambda)$,  $c_1, c_2, c_3$ are constant functions as opposed to the a priori possibility that they can be functions of the Gaussian null co-ordinates    $\Theta_i$, for all $k$-center solutions.
\item \textbf{(R3)} The appearance of a set of second order differential operators $\Delta^{(ij)}$ and the fact that each of the summands appearing in the first generalized Gegenbauer polynomial ${\cal G}_1$ are in the kernel of each of them, which implies $\Delta^{(ij)} ({\cal G}_1) = 0 $ for all $k$-center solutions.
\end{itemize}
We will see that similar reasons show up in the $d\geq 6$ multi center black hole and the multi center membrane solutions considered later.

So far, we have considered the most convenient way of characterizing the various $k$-center solutions. By convenient we mean the following. We chose a co-ordinate system for the transverse Euclidean space \eqref{spcoordinates}. For the two center solution, the line of black holes was conveniently chosen to be the $x_1$ axis. This makes the harmonic function a function of only one angle $\theta_1$. We had generalized Gegenbauer polynomial a function of the one angle $\theta_1$ and ${\cal G}_1$ consists of the first summand in \eqref{ggp1}. Then for the three center solution, the plane in which the black holes are in was conveniently chosen to be the span of the $x_1$ and $x_2$ axes. This makes the harmonic function a function of the first two angles $\theta_1$ and $\theta_2$. The generalized Gegenbauer polynomial is a function of $\theta_1$ and $\theta_2$  and ${\cal G}_1$ consists of the first two summands in \eqref{ggp1}. We had a convenient succession in that the $k$-center solution ($k \leq d-2$) had a dependence on the first $k-1$ angles $\theta_1, \ldots \theta_{k-1}$; the $k$-center generalized Gegenbauer polynomials are functions of these angles and the ${\cal G}_1$ consists of the first (k-1) summands in \eqref{ggp1}. And so on. But clearly one can think of inconvenient ways to characterize the various $k$-center solutions. Even for the two center solution, one can choose the line of black holes to be an arbitrary line, different from any of the $x_i$-axes. Then, even for the two center solution, one would have a dependence on all isotropic angles $\theta_i$ and the generalized Gegenbauer polynomials  would be functions of all isotropic angles and ${\cal G}_1$ would be consists of all the summands in \eqref{ggp1}. This inconvenient way of starting the problem can be done for any $k$-center solution. 

The results here in \ref{212} and \ref{213} can be seen as a solution to the $k$-center problem with the above inconvenient characterization. For this one only has to note that the charges $\mu_J$'s and co-ordinates $\vec{R}^{(J)}$ of all black holes  other than the first one are hidden in the definition of the generalized Gegenbauer polynomial and essentially disappear from view after one has  rewritten the harmonic function as \eqref{5dharm}. They show up in the final answer again because the final answer is in terms of the same generalized Gegenbauer polynomials. What this means is that the solution that has been obtained is true for all  possible values of charges and co-ordinates of the black holes; including values of the co-ordinates of the black holes in the two center problem (or any $k$-center problem) with the inconvenient characterization.  The solution to the two center problem (or any $k$-center solution) with the inconvenient characterization is thus identical to the solution of the most generic solution we have obtained here in \ref{212} and \ref{213}. This fact, that the solution here includes all possible values of co-ordinate positions of the black holes, is transparent only because of the use of generalized Gegenbauer polynomials; reformulating the starting point \eqref{5dharm} in terms of them and obtaining the final answers in terms of them.  One would like to think this also partially answers the question as to why the degree of horizon smoothness is identical for all $k$-center solutions. 

\subsubsection{\label{214} Solution in an alternate isotropic co-ordinate system}
In this section, we will consider an alternate isotropic co-ordinate system. An isotropic co-ordinate system is the $t$ co-ordinate that appears in \eqref{rnsoln} and any co-ordinate system for the transverse Euclidean space. Instead of \eqref{spcoordinates}, we will choose the following co-ordinates for the transverse $\mathbf{R}^4$, which basically amounts to choosing an alternate co-ordinate system for the three sphere. 
\be
x_1 = r  \sin \theta_1 \cos \theta_2, \quad x_2 = r \sin \theta_1 \, \sin \theta_2, \quad x_3 = r  \cos \theta_1 \cos \theta_3, \quad x_4 = r \cos \theta_1 \, \sin \theta_3, 
 \label{a5disotropic}
\ee
in which the flat metric takes the form
\be\label{a5dflatmetric}ds^2_{\mathbf{R}^{4}} = dr^2 + r^2 d\theta_1^2 + r^2\,\sin^2\theta_1\,d\theta_2^2   +  r^2\cos^2\theta_1\,  d\theta_3^2 . \ee
For the above alternate isotropic co-ordinate system, we are still using the co-ordinates $t, r, \theta_1, \theta_2, \theta_3$. The $t$ and $r$ here are the same as the previous isotropic co-ordinates with the same name, but the $\theta_i$ are clearly different. We choose to retain the same names so that we do not have to rewrite many of the formulae. The definition of the generalized Gegenbauer polynomials proceeds along the lines of \eqref{fR}-\eqref{ggp1} but with \eqref{fR} replaced by 
\ben \label{afR}
f^{(J)}(\theta_1, \theta_2, \theta_{3})  = \frac{R_1^{(J)}}{\|\vec{R}^{(J)} \|} \sin \theta_1 \cos \theta_2 +\frac{R_2^{(J)}}{\|\vec{R}^{(J)} \|} \sin \theta_1 \sin \theta_2 +\frac{R_3^{(J)}}{\|\vec{R}^{(J)} \|} \cos \theta_1 \cos \theta_3  +\frac{R_4^{(J)}}{\|\vec{R}^{(J)} \|} \cos \theta_1 \sin \theta_3 \nonumber \\\een 
and \eqref{ggp1} replaced by
\begin{multline} \label{aggp1} {\cal G}_1 (\theta_1, \theta_2, \theta_3) = \sum_{i=2}^{\infty} \frac{2 \mu_J}{\|\vec{R}^{(J)} \|^{4}}\,\left[R_1^{(J)}  \sin \theta_1 \cos \theta_2 +R_2^{(J)} \sin \theta_1 \sin \theta_2 +R_3^{(J)} \cos \theta_1 \cos \theta_3 \right. \\ \left. + R_4^{(J)} \cos \theta_1 \sin \theta_3\right].
\end{multline}
The harmonic function is \eqref{5dharm} but with the above defined ${\cal G}_n$'s. Note that there is no convenient way to choose the line of black holes in the two center solution so that the solution depends on only one angle. The best one can do is the solution depends on at least two angles, generically it depends on all angles. Unlike the previous isotropic co-ordinate system used in \ref{211}-\ref{213}, there is no convenient succession: two center depends on one angle, three center depends on two angles, $k$-center depends on $k-1$ angles etc.  We should think of any $k$-center solution described generically so that the solution and the generalized Gegenbauer polynomials depends on all angles and the first one ${\cal G}_1$ contains all summands. Different $k$'s correspond to different values for the co-ordinate positions $\vec{R}^{(i)}$ in \eqref{afR}.

To construct the Gaussian null co-ordinate system for the first horizon, we will follow the steps laid on in \ref{211}. The only changes to be made from there are that in the $\theta_i$-geodesic equation \eqref{5dthetaigeod} we now have
\be \label{5dFia} F_1(\theta_1, \theta_2, \theta_3) = 1, \quad F_2(\theta_1, \theta_2, \theta_3) =  \sin^2 \theta_1, \quad F_3(\theta_1, \theta_2, \theta_3) =  \cos^2 \theta_1 \ee
and the null condition \eqref{5dnull1} is replaced with 
\be \label{a5dnull}- H^{-2}\,\dot{t}^2 + H\,\dot{r}^2 + H r^2\,\dot{\theta_1}^2 + Hr^2\,\sin^2\theta_1\, \dot{\theta_2}^2 + Hr^2\,\cos^2\theta_1\,\dot{\theta_3}^2 = 0,\ee
after using \eqref{5dtgeode} becomes
\be \label{a5dnull1}- H + \dot{r}^2 +  r^2\,\dot{\theta_1}^2 + r^2\,\sin^2\theta_1\,
\dot{\theta_2}^2 + r^2\,\cos^2\theta_1\,\dot{\theta_3}^2 = 0.\ee
Note that the boundary conditions \eqref{5drbound}, \eqref{anglebound} and the final series expansion ansatz \eqref{expansionansatz} are unchanged. Then we ask the question, what is the minimal number of $c_n$'s and $b^{(i)}_n$'s needed to check for our expectations? Due to unchanged series expansion ansatze and the similar formula for the harmonic function, most of the fomulae from appendix \ref{A} are unchanged, except for \eqref{5dgtt33before}, \eqref{5dgtt13before}, \eqref{5dgtt23befor}. But the conclusion does not change. Hence we again have the reduced goal of solving the geodesic equations only upto the point needed to obtain $c_1 - c_6$ and $b^{(i)}_3 - b^{(i)}_5$ for each $i=1,2,3$.

Now we solve the null condition first \eqref{a5dnull1}. As before \ref{212}, to obtain $c_1-c_6$ we only need to consider the first two terms. Since the harmonic function here has the same form as before \eqref{5dharm}, we readily have the solution from the null condition: 
\begin{multline} \label{5drtfa} r(\lambda, v, \Theta_1, \Theta_2, \Theta_3)  = \sqrt{2}\mu_1^{1/4}\,\lambda^{1/2} + \frac{1}{2\sqrt{2}\mu_1^{1/4}}{\cal G}_0 \,\lambda^{3/2} 	+ \frac{2}{5}{\cal G}_1\,\lambda^2  -\frac{1}{48 \sqrt{2} \mu _1^{3/4}}\left[ 3\, {\cal G}_0^2 - 32\mu_1\, {\cal G}_2  \right]  \lambda^{5/2} 	\\  -\frac{2}{35 \mu _1^{1/2}} \left[ {\cal G}_0 \,{\cal G}_1  -10 \mu_1\,{\cal G}_3 \right] \lambda^3 + \ldots   \end{multline}
In all the results from  \eqref{5drtfa} onwards here in \ref{214}, the ${\cal G}_n$'s are the generalized Gegenbauer polynomials defined here (involving \eqref{afR} and are functions of $\Theta_1, \Theta_2, \Theta_3$. 
Now, we consider the $\theta_i$-geodesic equations. All considerations in \ref{212} hold again, there is a decoupling between the equations at least for the required coefficients and we obtain the solution:
\begin{multline} \label{5dthetaitfa}
\theta_i(\lambda, v, \Theta_1, \Theta_2, \Theta_3) = \Theta_i + \frac{\sqrt{2} }{\mu _1^{1/4}} \frac{\partial_{\Theta_i} {\cal G}_1}{F_i(\Theta_1, \Theta_2, \Theta_3)}\,\lambda^{3/2} + \frac34 \frac{\partial_{\Theta_i} {\cal G}_2}{F_i(\Theta_1, \Theta_2, \Theta_3)} \,\lambda^2 \\   -\frac{1}{10 \sqrt{2} \mu _1^{3/4}} \frac{17 \,{\cal G}_0\, \partial_{\Theta_i} {\cal G}_1 - 8 \mu _1\, \partial_{\Theta_i} {\cal G}_3 }{F_i(\Theta_1, \Theta_2, \Theta_3)} \lambda^{5/2} + \ldots
\end{multline} 

where $F_i(\Theta_1, \Theta_2, \Theta_3)$ are defined in \eqref{5dFia}.

After obtaining the definition of the Gaussian null co-ordinate system to the required order in \eqref{5drtfa}- \eqref{5dthetaitfa}, we can compute the tensor components. Most of the formulae are unchanged from \ref{213}. This has got to do with the facts that the series expansion ansatz\eqref{expansionansatz} is the same as before, the form of the harmonic function is the same as before \eqref{5dharm} and the solution to the required $c_n$'s is formally the same. $A_{\lambda}$ is given by \eqref{5dAl}, $A_v$ is given by \eqref{5dAv} and the $A_{\Theta_i}$ are given by \eqref{5dAThetai}.  $g_{vv}$ is given by \eqref{5dgvv}, $g_{v\Theta_i}$ are given by \eqref{5dgvThetai}, noting the crucial difference that all the generalized Gegenbauer polynomials appearing in these formulae are to be understood to be the ones defined here \ref{214}. The components $g_{\Theta_i \Theta_j}$ are also given by \eqref{5dgii} and \eqref{5dgij} but with a crucial difference. The $F_i(\Theta_1, \Theta_2, \Theta_3)$ are the ones in \eqref{5dFia} and the $\Delta^{(ij)}$'s are the following different six second order differential operators:
\ben \label{5dDeltaa}\Delta^{(11)} &=& 1+ \frac{\partial^2~~}{\partial \Theta_1^2},~~~\qquad \Delta^{(22)} = 1 + \cot \Theta_1 \frac{\partial~~ }{\partial \Theta_1} + \frac{1}{\sin^2 \Theta_1} \frac{\partial^2 ~~}{\partial \Theta_2}  \nonumber \\ 
\Delta^{(33)} &=&1-\tan \Theta_1 \frac{\partial}{\partial \Theta_1}  + \frac{1}{\cos^2 \Theta_1} \frac{\partial^2 ~~}{\partial \Theta_3^2}, \qquad ~~~\Delta^{(23)} = \frac{\partial^2 ~~}{\partial \Theta_2  \partial \Theta_3}  \nonumber \\ 
\Delta^{(12)} &=& \frac{\partial^2 ~~}{\partial \Theta_1  \partial \Theta_2} - \cot \Theta_1 \frac{\partial ~~}{\partial \Theta_2}, \quad \Delta^{(13)} = \frac{\partial^2 ~~}{\partial \Theta_1  \partial \Theta_3} + \tan \Theta_1 \frac{\partial ~~}{\partial \Theta_3}.   \een
We again have the remarkable fact  that each of the functions of angles  that appear as summands in \eqref{aggp1} (with the $\theta_i$'s replaced by the $\Theta_i$'s)  is in the kernel of each of the $\Delta^{(ij)}$'s. We again have 
\be \label{akernel} \Delta^{(ij)}({\cal G}_1) = 0, \qquad \text{for all $k$-center solutions.}\ee

 Thus, the result of the computations in this alternate isotropic co-ordinate system \eqref{a5disotropic} is also that: the horizon smoothness is identical for all $k$-center solutions including the most generic solution, the $\infty$-center solution and the degree of differentiability of individual tensor components (when it is finite) is identical for all $k$-center solutions. Again the underlying reasons are the same as the \textbf{(R1)}, \textbf{(R2)} and \textbf{(R3)} given in \ref{213}; but with a different set of $\Delta^{(ij)}$'s (\eqref{5dDeltaa} in place of \eqref{5dDeltao}) in \textbf{(R3)}.

We have worked with two different co-ordinate systems for the sphere in the transverse space, one here in \ref{214} the other in \ref{211}-\ref{213}, and obtained almost identical answers, for the transition functions and for the tensor components in the Gaussian null co-ordinate system, and identical conclusions for horizon smoothness and the degree of differentiability of individual components, \textbf{(R1)}, \textbf{(R2)} and \textbf{(R3)} .  This suggests that perhaps there is a way of solving the problem independent of choosing a particular co-ordinate system for the transverse sphere; perhaps the $\Delta^{(ij)}$'s appearing in \eqref{5dDeltao} and \eqref{5dDeltaa} are the same operators. We will not pursue this here, though.

\subsection{\label{22}$d \geq 6$}
We treat all dimensions bigger than five simultaneously. The procedure is identical to the $d=5$ case. We will be brief here. 
\subsubsection{\label{221}Constructing the Gaussian null co-ordinate system}
We need to solve the geodesic equations for the most generic solution. The solution to the $t$-geodesic equations is identical to the $d=5$ case. 
\ben \label{6dtgeode}\frac{d}{d\lambda} \left[ H^{-2} \, \frac{dt}{d\lambda} \right] = 0 ~&\Longrightarrow& ~ \frac{d}{d\lambda}t(\lambda) = -  H(r(\lambda), \theta_1(\lambda), \ldots \theta_{d-2}(\lambda))^2 \nonumber \\ ~ &\Longrightarrow& ~  t(\lambda) = v \,- \int d\lambda \,H(r(\lambda), \theta_1(\lambda), \ldots \theta_{d-2}(\lambda))^2, \een
Again $t(\lambda)$ is determined via \eqref{6dtgeode} in terms of $r(\lambda), \theta_1(\lambda), \ldots \theta_{d-2}(\lambda)$,  which are obtained by solving simultaneously the other geodesic  equations. We will solve the  ``$\theta_i$-geodesic'' equations for $i = 1, 2, \ldots d-2$,
\be \label{6dthetaigeod} \ddot{\theta_i} - H^{\frac{d-5}{d-3}}\,\frac{\partial_{\theta_i} H}{r^2 F_i(\theta_1, \ldots \theta_{d-2})} - \frac{\partial_{\theta_i} H}{(d-3) H r^2 F_i(\theta_1, \ldots \theta_{d-2})}\,\dot{r}^2  + \frac{2}{r}\,\dot{r} \,\dot{\theta_i} + \frac{2\,\partial_r H}{(d-3)H}\,\dot{r}\,\dot{\theta_i} + \ldots = 0,\ee
where
\begin{multline} \label{6dFi} F_1(\theta_1, \ldots \theta_{d-2}) = 1,\qquad F_2(\theta_1, \ldots \theta_{d-2}) = \sin^2 \theta_1, \qquad  F_3(\theta_1, \ldots \theta_{d-2}) = \sin^2 \theta_1 \sin^2 \theta_2, \ldots \\   F_{d-2}(\theta_1, \ldots \theta_{d-2}) = \sin^2 \theta_1 \sin^2 \theta_2, \ldots \sin^2 \theta_{d-3}\end{multline}
and the null condition
\be \label{6dnull}- H^{-2}\,\dot{t}^2 + H^{\frac{2}{d-3}}\,\dot{r}^2 + H^{\frac{2}{d-3}} r^2\,\left[\dot{\theta_1}^2 + \sin^2 \theta_1\,\dot{\theta_2}^2+ \ldots + \sin^2 \theta_1 \ldots \sin^2 \theta_{d-3}\,\dot{\theta}_{d-2}^2\right]  = 0\ee
which after using \eqref{6dtgeode} becomes
\be \label{6dnull1}- H^{\frac{2d-8}{d-3}} + \dot{r}^2 +  r^2\,\left[\dot{\theta_1}^2 + \sin^2 \theta_1\,\dot{\theta_2}^2+ \ldots + \sin^2 \theta_1 \ldots \sin^2 \theta_{d-3}\,\dot{\theta}_{d-2}^2\right]  = 0.\ee
The boundary conditions are identical to the $d=5$ case
\be \label{6dbound} r(\lambda = 0) = 0,\qquad \theta_i(\lambda = 0) = \Theta_i, \qquad \dot{\theta_i}(\lambda = 0) = 0,\qquad  i  = 1, 2, \ldots d-2.   \ee
Again we assume a series expansion for each of the unknown functions $r(\lambda)$, $\theta_i(\lambda)$. The expansion parameter is an appropriate power of the affine parameter $\lambda$ and is determined as before. Near the horizon, the leading (in $\lambda$) behavior of  the null condition:
\be \dot{r}^2 =  H^{\frac{2(d-4)}{d-3}} \quad \Longrightarrow \quad \dot{r}^{2} \sim \frac{1}{r^{2(d-4)}} \quad \Longrightarrow \quad  r(\lambda)^{d-3} \sim \lambda \qquad \Longrightarrow \qquad r(\lambda) \sim \lambda^{\frac{1}{d-3}}.\ee
Hence we assume the following series expansion ansatz : 
\be \label{264} r(\lambda) = \sum_{n = 0}^\infty c_n \, \left(\lambda^{\frac{1}{d-3}}\right)^n,\qquad \theta_i(\lambda) = \sum_{n = 0}^\infty b^{(i)}_n \, \left(\lambda^{\frac{1}{d-3}}\right)^n,\qquad  i = 1, 2, \ldots d-2\ee
The boundary conditions \eqref{5drbound}, \eqref{anglebound} imply the following coefficients vanish
\ben c_0 = 0, \quad  b^{(i)}_1 &=&  0,\quad b^{(i)}_2 = 0, \ldots \quad b^{(i)}_{d-3} = 0. \een 
We thus have 
\be \label{6dexpansionansatz} r(\lambda) = \sum_{n = 1}^\infty c_n \, \left(\lambda^{\frac{1}{d-3}}\right)^n,\qquad \theta_i(\lambda) = \Theta_i + \sum_{n = d-2}^\infty b^{(i)}_n \, \left(\lambda^{\frac{1}{d-3}}\right)^n.\ee
The procedure to obtain the solutions to the geodesic equations is to plug in the expansions \eqref{6dexpansionansatz} into the geodesic equations, obtain a series expansion of the equations in $\lambda^{\frac{1}{d-3}}$ and solve order by order. One would obtain the coefficients $c_n$'s and the $b^{(i)}_n$'s as functions of the constants $\Theta_i$. The solutions to the geodesic equations are hence  functions of the affine parameter $\lambda$ and the constants: $r( \lambda, \Theta_1, \ldots, \Theta_{d-2}), \theta_i(\lambda, \Theta_1, \ldots, \Theta_{d-2})$. One then uses \eqref{6dtgeode} to obtain 
\ben \label{269} t(\lambda)  &=& v \,- \int d\lambda \,H(r(\lambda), \theta_1(\lambda), \ldots \theta_{d-2}(\lambda))^2 \nonumber \\ &\equiv& v \,- T( \lambda, \Theta_1, \ldots, \Theta_{d-2}) \een
Before we implement this procedure, as before, we will ask ourselves the question: What is the minimal number of the $c_n$'s and the $b^{(i)}_n$'s  needed to check for the expectations one has for  the horizon smoothness of the most generic solution? 

We start with $g_{vv}$ whose tensor transformation law is given by $g_{vv} = -H^{-2}$ and using \eqref{6dexpansionansatz} we can see that its series expansion starts from $\lambda^2$. In the three center solution \cite{Gowdigere:2014aca}, it has non-zero coefficients only for $\lambda^2$, $\lambda^3$ and for $\lambda^{\frac{3d-8}{d-3}}$ terms, thus making it a $\mathcal{C}^3$ function. In going from three to four center etc we expect that $g_{v\Theta_i}$ follows $\textbf{(P3)}$. Hence we expect that $g_{vv}$ in the most generic solution will have a similar series expansion and to check this we will need to know the coefficients $c_1-c_{d-1}$.  Next, we consider $g_{v\Theta_i}$ whose tensor transformation law is given by $g_{v\Theta_i} = \frac{\partial_{\Theta_i} T}{H^2}$ and using \eqref{6dexpansionansatz} we can see that their series expansion starts from $\lambda^1$. In the three center solution \cite{Gowdigere:2014aca} (for $i=1,2$) they have non-zero coefficients only for $\lambda^1$, $\lambda^2$ and for $\lambda^{\frac{2d-5}{d-3}}$ terms, thus making them $\mathcal{C}^2$ functions. Hence we expect that $g_{v\Theta_i}$ for $i=1,2$, in the most generic solution will have a similar series expansion and to check this we will need to know the coefficients $c_1-c_{d-1}$. It turns out the knowing $c_1-c_{d-1}$ is suffiicient to check the expectiations even for $g_{v \Theta_i}$ $i >2$. Now, we consider $g_{\Theta_i \Theta_j}$ whose tensor transformation law is given by
\begin{multline} g_{\Theta_i \Theta_i} =  -\frac{\partial_{\Theta_i}T\,\partial_{\Theta_i}T}{H^2}  + \partial_{\Theta_i}r\,\partial_{\Theta_i}r\, H^{\frac{2}{d-3}} + \partial_{\Theta_i}\theta_1\,\partial_{\Theta_i}\theta_1\, H^{\frac{2}{d-3}} \,r^2 \\ + \partial_{\Theta_i}\theta_2\,\partial_{\Theta_i}\theta_2\, H^{\frac{2}{d-3}} \,r^2\,\sin \theta_1^2 + \ldots + \partial_{\Theta_i}\theta_{d-2}\,\partial_{\Theta_i}\theta_{d-2}\, H^{\frac{2}{d-3}} \,r^2\,\sin \theta_1^2 \ldots \sin^2 \theta_{d-3}\end{multline}
and using \eqref{6dexpansionansatz} we can see that their series expansion starts from $\lambda^0$. In the three center solution \cite{Gowdigere:2014aca} (for $i=1,2$) the earliest fractional order is $\lambda^{\frac{d-1}{d-3}}$, thus making them $\mathcal{C}^1$ functions. Hence we expect that $g_{\Theta_i \Theta_j}$ in the most generic solution will have a similar series expansion and to check this we will need to know the coefficients $c_1-c_d$ and $b^{(i)}_{d-2}, b^{(i)}_{d-1}$ (for all $i$),  which is also enough, it turns out to check the expectiations  for $g_{\Theta_i \Theta_j}$ with $i,j >2$. For $A_{\lambda}$, the tensor transformation law is given by $A_\lambda = H$ and using \eqref{6dexpansionansatz} we can see that its series expansion starts from $\lambda^{-1}$. In the three center solution \cite{Gowdigere:2014aca}, it has non-zero coefficients only for $\lambda^0$  and for $\lambda^{\frac{d-2}{d-3}}$ terms $\lambda^0$ (apart from a pure gauge term at $\lambda^{-1}$) thus making it a $\mathcal{C}^0$ function. In going from three to four center etc we expect that $A_\lambda$ follows $\textbf{(P3)}$. Hence we expect that $A_{\lambda}$ in the most generic solution will have a similar series expansion and to check this we will need to know the coefficients $c_1-c_{d-1}$. For $A_{v}$, the tensor transformation law is given by $A_v  = - H^{-1}$ and using \eqref{6dexpansionansatz} we can see that its series expansion starts from $\lambda^1$. In the three center solution \cite{Gowdigere:2014aca}, it has non-zero coefficients only for $\lambda^1$, $\lambda^2$  and for $\lambda^{\frac{2d-5}{d-3}}$ terms, thus making it a $\mathcal{C}^2$ function. In going from three to four center etc we expect that $A_{v}$ follows $\textbf{(P3)}$. Hence we expect that $A_{v}$ in the most generic solution will have a similar series expansion and to check this we will need to know the coefficients $c_1-c_{d-1}$.  For $A_{\Theta_i}$, the tensor transformation law is given by $A_{\Theta_i} = \frac{\partial_{\Theta_i}T}{H}$ and using \eqref{6dexpansionansatz} we can see that its series expansion starts from $\lambda^0$. In the three center solution \cite{Gowdigere:2014aca}, $A_{\Theta_i}$ for $i =1,2$ has non-zero coefficients only for $\lambda^0$, $\lambda^1$  and for $\lambda^{\frac{d-2}{d-3}}$ terms, thus making it a $\mathcal{C}^1$ function. In going from three to four center etc we expect that $A_{\Theta_i}$ follows $\textbf{(P3)}$. Hence we expect that $A_{\Theta_i}, i =1,2$ in the most generic solution will have a similar series expansion and to check this we will need to know the coefficients $c_1-c_{d-1}$; which it turns out is sufficient to check for the expectations of $A_{\Theta_i}, i >2$ as well.

To conclude, we set ourselves the much reduced goal of solving the geodesic equations only upto the point needed to obtain $c_1 - c_d$ and $b^{(i)}_{d-2},  b^{(i)}_{d-1}$ for each $i=1, 2, \ldots d-2$.

\subsubsection{\label{222} Solving the geodesic equations}
We now solve the geodesic equations. It is convenient to solve the $\theta_i$-geodesic equations \eqref{6dthetaigeod} together with the null condition \eqref{6dnull1}. As it happened for $d=5$, we will see that there is a decoupling of sorts that happens: the coefficients $c_1-c_d$ are determined by the null condition, the coefficients $b^{(i)}_{d-2}, b^{(i)}_{d-1}$ for any $i$ are determined by the $\theta_i$-geodesic equation for that $i$.

\textbf{Null condition:} Using  \eqref{6dexpansionansatz}, we can work out expansion of the (left hand side of the) null condition \eqref{6dnull1}. The last $d-2$ terms start at order four while the first two terms start at order $-(2d-8)$. Hence the first $2d-4$ non-trivial orders of the null condition, which are the orders from $-(2d-8)$ to plus three, receive contributions from only the first two terms. Using \eqref{6dexpansionansatz} carefully, one can, in a manner similar to the $d=5$ analysis of \ref{212}, show that the coefficients in the first $2d-4$ non-trivial orders of the null condition are functions of only the $c_1 - c_{2d-4}$ with none of the $b^{(i)}_n$'s making an appearance. We need only a subset of them, $c_1-c_d$ which we readily obtain:
\begin{multline} \label{6drsoln} r(\lambda, v, \Theta_1, \ldots, \Theta_{d-2}) = (d-3)^{1/{d-3}}\, \mu_1^{\frac{d-4}{(d-3)^2}}\, \lambda^{1/{d-3}} +\frac{d-4}{2d-6}\,(d-3)^{\frac{1}{d-3}}\, \mu_1^{-\frac{1}{(d-3)^2}}\,{\cal G}_{0} \,  \lambda^{\frac{d-2}{d-3}}  \\  + \frac{d-4}{2d-5}\,(d-3)^{\frac{2}{d-3}}\, \mu_1^{\frac{d-5}{(d-3)^2}}\,{\cal G}_{1} \,   \lambda^{\frac{d-1}{d-3}}  + \frac{d-4}{2d-4}\,(d-3)^{\frac{3}{d-3}}\, \mu_1^{\frac{2 d-9}{(d-3)^2}}\,{\cal G}_{2} \,  \lambda^{\frac{d}{d-3}} + \ldots \end{multline}
Note that 
\ben  \label{vanishingc} c_2 =  0,\qquad c_3 =  0, \ldots \ldots \qquad c_{d-3} = 0. \een 
In the above, and in every formula in \ref{222}, ${\cal G}_n$'s appearing are all functions of the Gaussian null co-ordinate angles $\Theta_1, \ldots \Theta_{d-2}$.

Similar to what happened for the $d=5$ case, for any $k$-center solution, the result for $c_1 - c_d$ will still be given by \eqref{6drsoln}, with the understanding that one has to replace with generalized Gegenbauer polynomials appropriate for $k$-center solution, i.e. the ones with $k-1$ summands in \eqref{fR}. Hence the result \eqref{6drsoln} for \emph{ $c_1 - c_d$ is independent of $k$ (the number of arbitrarily positioned centers).} This feature has it's origin in the fact that up to this order in the computation none of the $b^{(i)}_n$'s show up. The $b^{(i)}_n$'s are accompanied by derivatives of the generalized Gegenbauer polynomials which will be different for different $k$.  This independence from $k$ of the results of $c_1 - c_d$ will feature in subsequent analysis. 

\textbf{$\theta_i$-geodesic equations :} We begin by working out the series expansion of the $\theta_i$-geodesic equations.  The terms that we have not displayed in \eqref{6dthetaigeod} are the ones are proportional to $\dot{\theta_j}\,\dot{\theta_k}$ and start from order two.  The terms displayed are the ones that contribute to the first $d-2$ orders from order $-(d-4)$ to plus one.  Evaluating these orders using \eqref{6dexpansionansatz} shows that they are functions only of (i) the already determined coefficients $c_n$'s \eqref{6drsoln} and of (ii) $b^{(i)}_{d-2}, \ldots$, with none of the $b^{(j)}_n$'s for $j \neq i$ making an appearance (however, they do make an appearance from order two onwards).  Thus we have a decoupling similar to the $d=5$ case: for a given $i$, the required coefficients $b^{(i)}_{d-2}, b^{(i)}_{d-1}$ appear (earliest in the series expansion) only in the $\theta_i$-geodesic equation for that $i$. By solving only the first two orders of the $\theta_i$-geodesic equations, we obtain the required coefficients, for $i = 1, 2, \ldots d-2$: 
\begin{multline} \label{6dthetaisoln} \theta_i(\lambda)  = \Theta_i +  (d-3)^{\frac{1}{d-3}}\,\mu_1^{-\frac{1}{(d-3)^2}}\,\frac{\partial_{\Theta_i} {\cal G}_{1}}{F_i(\Theta_1,\ldots, \Theta_{d-2})} \, \lambda^{\frac{d-2}{d-3}} +\\ \frac{1}{2}\,(d-3)^{\frac{2}{d-3}}\,\mu_1^{\frac{d-5}{(d-3)^2}}\,\frac{\partial_{\Theta_i} {\cal G}_{2}}{F_i(\Theta_1,\ldots, \Theta_{d-2})}\,\lambda^{\frac{d-1}{d-3}}  +\ldots \end{multline}
where $F_i(\Theta_1,\ldots, \Theta_{d-2})$ are given in \eqref{6dFi}. 

Using the above, one can compute \eqref{269} and obtain
\be \label{6dttf} t( \lambda, v, \Theta_1, \ldots \Theta_{d-2}) = v - \, T(\lambda, v, \Theta_1, \ldots \Theta_{d-2}) ,\ee
where 
\begin{multline} \label{6dT}
T(\lambda, \Theta_1, \ldots, \Theta_{d-2}) =  - \frac{1}{(d-3)^2}\,\mu _1^{2/{d-3}}\, \lambda^{-1} \left[1  -     (d -2)\,\mu _1^{-1/{d-3}}\,{\cal G}_0 \, \lambda \,\log \lambda \right. \\ \left.   -    \frac{2 d -2}{2 d - 5}\,(d-3)^{\frac{2d-5}{d-3}} \,\mu_1^{-\frac{1}{(d-3)^2}}\,{\cal G}_1 \, \lambda^{\frac{d -2}{d-3}}  -    \frac{d}{2 d-4}\,(d-3)^{\frac{2d-4}{d-3}} \,\mu_1^{\frac{d-5}{(d-3)^2}}\,{\cal G}_2 \, \lambda^{\frac{d -1}{d-3}} + \ldots \right]
\end{multline}
We have now obtained in \eqref{6drsoln}, \eqref{6dthetaisoln} and in \eqref{6dttf} the minimally needed definition of the horizon co-ordinate system.

\subsubsection{\label{223}Tensor components in Gaussian null co-ordinates}
To obtain the series expansions of the components of the metric and gauge fields in the Gaussian null co-ordinate system for the most generic solution, we plug in the transition functions obtained in \ref{222} into the tensor transformation laws. Again we do not need to compute the components given in \eqref{lambdarow}. The other metric components are given by:
\begin{multline} \label{6dgvv} g_{vv} = - (d-3)^2 \mu _1^{-2/{d-3}}\,\lambda^2 +  (d-2)(d-3)^2\, \mu _1^{-3/{d-3}}\, {\cal G}_0\,\lambda^3  \\ + \frac{2d-2}{2d-5}\,(d-3)^{\frac{3d-8}{d-3}}\,\mu_1^{\frac{2d-5}{(d-3)^2}}{\cal G}_1 \,\lambda^{\frac{3d-8}{d-3}} +\dots  \end{multline}
\ben  \label{6dgvt} g_{v\Theta_i}  &=&  \frac{2 d - 2}{2 d-5}\, (d-3)^{\frac{2 d-5}{d-3}} \, \mu _1^{-\frac{1}{(d-3)^2}} \partial_{\Theta_i} {\cal G}_1 \,\lambda ^{\frac{2 d-5}{d-3}} + \ldots   \een
\begin{multline} \label{6dgtt} \frac{g_{\Theta_i\Theta_i}}{F_i(\Theta_1, \ldots, \Theta_{d-2})} =  \mu _1^{\frac{2}{d-3}} + 2 \, \mu _1^{\frac{1}{d-3}} {\cal G}_0\,\lambda  + 2 (d-3)^{\frac{1}{d-3}} \mu_1^{\frac{2d-7}{(d-3)^2}}\Delta^{(ii)}({\cal G}_1)\,\lambda^{\frac{d-2}{d-3}} \\ + \frac{d-2}{d-1} (d-3)^{\frac{2}{d-3}} \mu _1^{\frac{3 d-11}{(d-3)^2}} \left(\Delta^{(ii)}\,+\frac{d}{d-2}\right) ({\cal G}_2) ~\lambda^{\frac{d-1}{d-3}}  + \ldots \end{multline}
\begin{equation} \label{6dgtp} g_{\Theta_i \Theta_j}  = 2 (d-3)^{\frac{1}{d-3}} \mu_1^{\frac{2d-7}{(d-3)^2}} \Delta^{(ij)}({\cal G}_1)\,\lambda^{\frac{d-2}{d-3}} +  \frac{d-2}{d-1}\,(d-3)^{\frac{2}{d-3}} \mu _1^{\frac{3 d-11}{(d-3)^2}}  \Delta^{(ij)}({\cal G}_2) ~\lambda^{\frac{d-1}{d-3}}  + \ldots \end{equation}
In the above $F_i(\Theta_1, \ldots \Theta_{d-2})$ is defined in \eqref{6dFi} and the $\Delta^{(ij)}$'s are second order differential operators, $\frac{(d-2)(d-1)}{2}$ of them, that have appeared already in \eqref{5dDeltao}.

The components of the gauge field are given by:
\ben \label{6dAlambda} A_\lambda = \frac{\mu _1^{\frac{1}{d-3}}}{(d-3)}\lambda^{-1} +  \frac{d-2}{2 (d-3)}{\cal G}_0 + \frac{d-1}{2 d-5}(d-3)^{\frac{1}{d-3}}\mu _1^{\frac{d-4}{(d-3)^2}}{\cal G}_1\lambda ^{\frac{1}{d-3}}  + \ldots \een 
\begin{multline} \label{6dAv} A_v = -(d-3) \mu _1^{-\frac{1}{d-3}} \lambda + \frac{1}{2} (d-2) (d-3)\mu _1^{-\frac{2}{d-3}}{\cal G}_0\,\lambda^2 +  \frac{d-1}{2 d-5}\,(d-3)^{\frac{2 d-5}{d-3}}\mu _1^{-\frac{d-2}{(d-3)^2}}{\cal G}_1\lambda ^{\frac{2 d-5}{d-3}}  + \ldots  \end{multline}
\ben \label{6dAt} A_{\Theta_i} = \frac{2 (d-1)}{2 d-5}\,(d-3)^{\frac{d-2}{d-3}}\,\mu _1^{\frac{d-4}{(d-3)^2}}\,\partial_{\Theta_i} {\cal G}_1\,\lambda ^{\frac{d-2}{d-3}} + \ldots.    \een
A perusal of the formulae we have obtained shows that all  the expectations we had for each of the components  are played out. All the comments we made for the $d=5$ case, between formulae \ref{5dgvv} and \ref{5dDeltao} hold here with the obvious changes. Again we have the remarkable fact that each of the $d-1$ summand functions of angles that occur in ${\cal G}_1$ \eqref{ggp1} are in the kernel of each of the  $\frac{(d-2)(d-1)}{2}$ differential operators \eqref{5dDeltao}. Since ${\cal G}_1$  for different $k$-center solutions is a (different) linear combination of these functions of angles, ${\cal G}_1$  is in the kernel of each of the  $\Delta^{(ij)}$'s for all $k$-center solutions. Thus we again have 
\be \label{6dkernel} \Delta^{(ij)}({\cal G}_1) = 0, \qquad \text{for all $k$-center solutions.}\ee

With the aid of the actual computations of tensor components in the Gaussian null co-ordinate system, we are able to see that the surmise we made for the $k+1$-center horizon smoothness to the $k$-center horizon smoothness and the consequent expectations are all realized in reality. We have thus shown  that \emph{the horizon smoothness is identical for all $k$-center solutions including the most generic solution, the $\infty$-center solution}.  Not only is the horizon smoothness identical for all $k$-center solutions, we have seen that even an  individual component (whenever it has a finite degree of differentiability) has identical series expansions and hence identical degree of differentiability for all $k$-center solutions (for which it has a finite degree of differentiability). Again we gather the underlying reasons behind this:
\begin{itemize}
\item \textbf{(R1)} The boundary conditions that determine the series ansatze for $r(\lambda)$, $\theta_i(\lambda)$ are identical for all $k$-center solutions. 
\item \textbf{(R2)} In the solution to the geodesic equations, the first $d-2$  coefficients in the series expansion for $r(\lambda)$,  $c_1, c_2, \ldots c_{d-2}$ are constant functions as opposed to the a priori possibility that they can be functions of the Gaussian null co-ordinates    $\Theta_i$, for all $k$-center solutions.
\item \textbf{(R3)} The appearance of a set of second order differential operators $\Delta^{(ij)}$ and the fact that each of the summands appearing in the first generalized Gegenbauer polynomial ${\cal G}_1$ are in the kernel of each of them, which implies $\Delta^{(ij)} ({\cal G}_1) = 0 $ for all $k$-center solutions.
\end{itemize}

\section{\label{3} The most generic multi center $M2$ branes}

The multi center $M2$ brane solutions we investigate are (bosonic) solutions to eleven dimensional supergravity. Following is the solution in  isotropic co-ordinates:
\be \label{m2soln}
ds^2 =  H^{-2/3}\,( -dt^2 + dx^2 + dy^2) + H^{1/3}\, ds^2_{\mathbf{R}^8}, \qquad  C_3 = \frac{dt}{H}\ee
where $ds^2_{\mathbf{R}^8}$ is the flat metric of the transverse Euclidean space $\mathbf{R}^8$. $H$ is a harmonic function in the transverse Euclidean space:
\be \label{m2harm}H(\vec{r}) = 1 + \sum_{J = 1}^\infty \frac{\mu_J}{\| \vec{r} - \vec{r}_J \|^{6}}\ee 
We will first introduce the co-ordinate system for the transverse Euclidean space given in \eqref{spcoordinates}, \eqref{flatmetric} with the substitution $d=9$. Later, in \ref{307}, we will consider a different co-ordinate system.  Thus, the co-ordinates in the isotropic co-ordinate system are $t, x, y, r, \theta_1, \ldots \theta_{7}$. The harmonic function for the most generic multi center $M2$-brane solution is given in \eqref{3cenharm3} with $d = 9$.

\subsubsection{\label{304}Constructing the horizon co-ordinate system}
The horizon co-ordinate system for the membrane horizon was worked out in \cite{Gowdigere:2012kq}. It is similar to the Gaussian null co-ordinate system in that it is constructed out of the radial null geodesics: the solutions to the geodesic equations provide transition functions to a co-ordinate system which covers the horizon. Hence we consider the solution to geodesic equations. 

$\frac{\partial}{\partial t}, \frac{\partial}{\partial x}$ and $\frac{\partial}{\partial y}$ are Killing vector fields of the metric, due to which the  $t$-geodesic, $x$-geodesic and $y$-geodesic equations can be integrated once: 
\be \label{11dtxygeod} \frac{d}{d\lambda} \left[ H^{-2/3} \, \frac{dt}{d\lambda} \right] = 0, \qquad \frac{d}{d\lambda} \left[ H^{-2/3} \, \frac{dx}{d\lambda} \right] = 0, \qquad \frac{d}{d\lambda} \left[ H^{-2/3} \, \frac{dy}{d\lambda} \right] = 0. \ee  
We will solve \eqref{11dtxygeod} in the following way \cite{Gowdigere:2012kq},
\ben \label{txysoln} t(\lambda) &=& v - f(v, X, Y)\,\int d\lambda \, H(r(\lambda), \, \theta_1(\lambda), \ldots  \theta_7(\lambda))^{2/3}, \nonumber \\  x(\lambda) &=& X - g(v, X, Y)\,\int d\lambda \, H(r(\lambda), \, \theta_1(\lambda), \ldots  \theta_7(\lambda))^{2/3}, \nonumber  \\ y(\lambda) &=& Y - h(v, X, Y)\,\int d\lambda \, H(r(\lambda), \, \theta_1(\lambda), \ldots  \theta_7(\lambda))^{2/3}, \een
where $f, g$ and $h$ are arbitrary smooth functions of the  integrations constants $v, X$ and $Y$. We chose to introduce the arbitrary smooth functions $f,g,h$ of integration constants in the above manner because a simple choices such as constant functions or all of them functions of one variable only,  won't provide a good horizon co-ordinate system.  It turns out that a completely arbitrary choice of functions $f,g,h$ does not work either. They will need to satisfy various conditions (see \cite{Gowdigere:2014aca} for all details). that we will encounter along the way. Although we do not have a solution to all the constraints that the $f,g,h$ would need to satisfy by the end of the analysis, we do have many examples:
\begin{eqnarray} \label{fghexamples}
f(v, X, Y) &=&\frac{1}{2}\left( X + \frac{1}{X} + \frac{Y^2}{X} \right), \quad g(v, X, Y)= \frac{1}{2}\left( - X + \frac{1}{X} + \frac{Y^2}{X} \right), \quad h(v, X, Y)= Y. \nonumber \\
f(v, X, Y)&=&\sqrt{1+ Y^2}\,\cosh X ,\quad g(v, X, Y)= \sqrt{1 + Y^2}\,\sinh X , \quad h(v, X, Y) = Y.
\end{eqnarray}
For the other functions, we solve the $\theta_i$-geodesic equations
\begin{multline} \label{11dthetaigeod} \ddot{\theta} + \frac{\partial_{\theta_i} H}{3 H^{2/3}r^2 F_i(\theta_1, \ldots \theta_7)} (-f^2 + g^2 + h^2) - \frac{\partial_{\theta_i} H}{6 H r^2 F_i(\theta_1, \ldots \theta_7)}\,\dot{r}^2 + \frac{2}{r}\,\dot{r} \,\dot{\theta_i} + \frac{\partial_r H}{3 H}\,\dot{r}\,\dot{\theta_i} + \ldots = 0.\end{multline}
where 
\begin{multline} \label{11dFi} F_1(\theta_1, \ldots \theta_{7}) = 1,\qquad F_2(\theta_1, \ldots \theta_{7}) = \sin^2 \theta_1, \qquad  F_3(\theta_1, \ldots \theta_{7}) = \sin^2 \theta_1 \sin^2 \theta_2,\qquad  \ldots \\\ldots  \qquad F_{7}(\theta_1, \ldots \theta_{7}) = \sin^2 \theta_1 \sin^2 \theta_2, \ldots \sin^2 \theta_{6}\end{multline}
and the null-condition
\be \label{11dnull} H^{-2/3}\,(-\dot{t}^2 + \dot{x}^2 + \dot{y}^2) + H^{\frac{1}{3}}\,\dot{r}^2 + H^{\frac{1}{3}}\, r^2\,\left[\dot{\theta_1}^2 + \sin^2 \theta_1\,\dot{\theta_2}^2+ \ldots + \sin^2 \theta_1 \ldots \sin^2 \theta_{6}\,\dot{\theta}_{7}^2\right]  = 0\ee
which after using \eqref{txysoln} becomes
\be \label{11dnull1} H^{1/3}\,(- f^2 + g^2 + h^2 ) + \dot{r}^2 +  r^2\left[\dot{\theta_1}^2 + \sin^2 \theta_1\,\dot{\theta_2}^2+ \ldots + \sin^2 \theta_1 \ldots \sin^2 \theta_{6}\,\dot{\theta}_{7}^2\right] = 0.\ee
We can now use one of the freedoms in defining the affine parameter to set 
\be \label{fghconstraint1} S ~\equiv~ -f^2 + g^2 + h^2 = -1. \ee
The boundary conditions are as before:
\be \label{11dbound} r(\lambda = 0) = 0,\qquad \theta_i(\lambda = 0) = \Theta_i, \qquad \dot{\theta_i}(\lambda = 0) = 0,\qquad  i  = 1, 2, \ldots 7.   \ee
Again we assume a series expansion for each of the unknown functions $r(\lambda)$, $\theta_i(\lambda)$. The expansion parameter is an appropriate power of the affine parameter $\lambda$ and is determined as before. Near the horizon, the leading (in $\lambda$) behavior of  the null condition:
\be \dot{r}^2 =  H^{1/3} \quad \Longrightarrow \quad \dot{r}^{2} \sim \frac{1}{r^{2}} \quad \Longrightarrow \quad  r(\lambda)^2 \sim \lambda \qquad \Longrightarrow \qquad r(\lambda) \sim \sqrt{\lambda}.\ee
Hence we assume the following series expansion ansatz : 
\be \label{391} r(\lambda) = \sum_{n = 0}^\infty c_n \, \left(\sqrt{\lambda}\right)^n,\qquad \theta_i(\lambda) = \Theta_i +  \sum_{n = 0}^\infty b^{(i)}_n \, \left(\sqrt{\lambda}\right)^n,\qquad  i = 1, 2, \ldots 7\ee
The boundary conditions \eqref{11dbound} imply the following coefficients vanish
\ben c_0 = 0, \quad  b^{(i)}_1 &=&  0,\quad b^{(i)}_2 = 0. \een 
We thus have 
\be \label{11dexpansionansatz} r(\lambda) = \sum_{n = 1}^\infty c_n \, \left(\sqrt{\lambda}\right)^n,\qquad \theta_i(\lambda) = \sum_{n = 3}^\infty b^{(i)}_n \, \left(\sqrt{\lambda}\right)^n, \quad i = 1,2, \ldots 7.\ee
The procedure to obtain the solutions to the geodesic equations is to plug in the expansions \eqref{11dexpansionansatz} into the geodesic equations, obtain a series expansion and solve order by order. One would obtain the coefficients $c_n$'s and the $b^{(i)}_n$'s as functions of the constants $\Theta_i$. The solutions to the geodesic equations are hence  functions of the affine parameter $\lambda$ and the constants: $r( \lambda, X, Y, \Theta_1, \ldots, \Theta_{7}), \theta_i(\lambda, X, Y, \Theta_1, \ldots, \Theta_{7})$.  We then get  from \eqref{txysoln}
\ben \label{11dtxytf} t(\lambda, X, Y, \Theta_1, \ldots, \Theta_{7}) &=& v - f(v, X, Y)\,T(\lambda, \Theta_1, \ldots, \Theta_{7}), \nonumber \\ x(\lambda, X, Y,\Theta_1, \ldots, \Theta_{7}) &=& X - g(v, X, Y)\,T(\lambda, \Theta_1, \ldots, \Theta_{7}), \nonumber \\ ~ y(\lambda, X, Y, \Theta_1, \ldots, \Theta_{7}) &=& Y -  h(v, X, Y)\,T(\lambda, \Theta_1, \ldots, \Theta_{7}) \een
where 
\be \label{11dT} T(\lambda, \Theta_1, \ldots, \Theta_{7}) \equiv \int d\lambda \, H(r(\lambda), \, \theta_1(\lambda), \ldots \theta_7(\lambda))^{2/3}.\ee

Before we implement this procedure, as before, we will ask ourselves the question: What is the minimal number of the $c_n$'s and the $b^{(i)}_n$'s  needed to check for the expectations one has for  the horizon smoothness of the most generic multi center $M2$ solution? We follow the steps that have already been implemented for the black hole case in \ref{211} and \ref{221} and find that we have the much reduced task of solving the geodesic equations only upto the point needed to obtain $c_1 - c_8$ and $b^{(i)}_3 - b^{(i)}_9$ for each $i=1,2, \ldots 7$.

\subsubsection{\label{305}Solving the geodesic equations}

We now solve the geodesic equations. It is convenient to solve the $\theta_i$-geodesic equations \eqref{11dthetaigeod} together with the null condition \eqref{11dnull1}. As it happened for black holes,  we will see that there is a decoupling of sorts that happens: the coefficients $c_1-c_8$ are determined by the null condition, the coefficients $b^{(i)}_{3}- b^{(i)}_{9}$ for any $i$ are determined by the $\theta_i$-geodesic equation for that $i$. Let us reiterate that in the rest of this section, i.e. \ref{305} and \ref{306}, all the generalized Gegenbauer polynomials that will be encountered are functions of the horizon co-ordinates ${\cal G}_n(\Theta_1, \ldots \Theta_7)$.

\textbf{Null condition:} We again repeat the steps as in \ref{212} and \ref{222}. The important features of that computations repeat themselves here and we obtain:
\be \label{11drtf} r(\lambda, X, Y, \Theta_1 \ldots \Theta_7) = \sqrt{2}\mu_1^{1/12} \lambda^{1/2} + \frac{1}{3 \sqrt{2} \mu _1^{5/12}} {\cal G}_0 \lambda^{7/2} + \frac{8}{27 \mu _1^{1/3}}  {\cal G}_1\lambda^4\ + \ldots\ee

\textbf{$\theta_i$-geodesic equations :} We solve as before and obtain:
\begin{multline}  \label{11dthetaitf} \theta_i(\lambda, X, Y, \Theta_1 \ldots \Theta_7) = \Theta_i + \frac{4 \sqrt{2}}{35 \mu _1^{5/12}} \frac{\partial_{\Theta_i}{\cal G}_1}{F_i(\Theta_1, \ldots \Theta_{7})}\, \lambda^{7/2} +  \frac{1}{6 \mu _1^{1/3}} \frac{\partial_{\Theta_i}{\cal G}_2}{F_i(\Theta_1, \ldots \Theta_{7})}\,\lambda^4\ \\ + \frac{8 \sqrt{2}}{63 \mu _1^{1/4}} \frac{\partial_{\Theta_i}{\cal G}_3}{F_i(\Theta_1, \ldots \Theta_{7})}\, \lambda^{9/2}  + \ldots\end{multline}
where the $F_i(\Theta_1, \ldots \Theta_7)$ are defined in \eqref{11dFi}. Using the above in \eqref{11dT} we have
\begin{multline} \label{Tm}
T(\lambda, \Theta_1, \Theta_2, \ldots \Theta_7) = -\frac{\mu_1^{1/3}}{4} \lambda^{-1} + \frac{7}{12 \mu_1^{1/6}}{\cal G}_ 0\,\lambda^2 + \frac{64 \sqrt{2}}{135 \mu_1^{1/12}}{\cal G}_1\,\lambda^{5/2} + \frac{4}{5}{\cal G}_ 2\,\lambda^3 + \frac{160\sqrt{2} \mu_1^{1/12}}{231}  {\cal G}_3 \lambda^{7/2}+\ldots
\end{multline}
In \eqref{11dtxytf}, \eqref{11drtf} and \eqref{11dthetaitf}, we have obtained the minimally needed definiton of the horizon co-ordinate system. 

\subsubsection{\label{306} Tensor components in horizon co-ordinates}

To obtain the series expansions of the components of the metric and gauge fields in the Gaussian null co-ordinate system for the most generic solution, we plug in the transition functions obtained in \ref{305} into the tensor transformation laws. Again we do not need to compute some of the metric components \cite{Gowdigere:2014aca}:
\be \label{11dlambdarow} g_{\lambda \lambda} = 0, \quad g_{\lambda v} = f, \quad g_{\lambda X} = -g, \quad g_{\lambda Y} = - h, \quad g_{\lambda \Theta_i} = 0. \ee
Hence the above metric components are smooth. The other metric components are given by:
\ben \label{11dgvv} g_{vv} &=&  \frac{1}{4} \mu _1^{1/3}  z_1 - 2\,  \partial_vf\,\lambda - \frac{4}{\mu _1^{1/3}} \lambda^2 -\frac{7}{3 \mu _1^{1/6}}  z_1 {\cal G}_0\, \lambda^3 -\frac{32 \sqrt{2} }{15 \mu _1^{1/{12}}} z_1 {\cal G}_1 \lambda^{7/2}  + \ldots \nonumber 
\\
 \label{11dgXX} g_{XX} &=&  \frac{1}{4} \mu _1^{1/3}  z_2 + 2\,  \partial_Xg\,\lambda + \frac{4}{\mu _1^{1/3}}\lambda^2 -\frac{7}{3 \mu _1^{1/6}}  z_2  {\cal G}_0\, \lambda^3 -\frac{32 \sqrt{2} }{15 \mu _1^{1/{12}}} z_2 {\cal G}_1 \lambda^{7/2}  + \ldots \nonumber \\
\label{11dgYY} g_{YY} &=&  \frac{1}{4} \mu _1^{1/3}  z_3 + 2\,  \partial_Yh\,\lambda + \frac{4}{\mu _1^{1/3}}\,\lambda^2 -\frac{7}{3 \mu _1^{1/6}}  z_3\,{\cal G}_0\, \lambda^3 -\frac{32 \sqrt{2} }{15 \mu _1^{1/{12}}} z_3\, {\cal G}_1 \lambda^{7/2}  + \ldots \nonumber \\
 \label{11dgvX} g_{vX} &=&  \frac{1}{4} \mu _1^{1/3}  q_2 -  q_1\, \lambda -\frac{7}{3 \mu _1^{1/6}}  q_2\, {\cal G}_0\, \lambda^3 -\frac{32 \sqrt{2} }{15 \mu _1^{1/{12}}} q_2\, {\cal G}_1 \lambda^{7/2}  + \ldots \nonumber \\ 
 \label{11dgvY} g_{vY} &=&  \frac{1}{4} \mu _1^{1/3}  q_4 -  q_3\, \lambda -\frac{7}{3 \mu _1^{1/6}}  q_4 \, {\cal G}_0\, \lambda^3 -\frac{32 \sqrt{2} }{15 \mu _1^{1/{12}}} q_4\, {\cal G}_1 \lambda^{7/2}  + \ldots \nonumber \\
 \label{11dgXY} g_{XY} &=&  \frac{1}{4} \mu _1^{1/3}  q_6  -  q_5\, \lambda -\frac{7}{3 \mu _1^{1/6}}  q_6 \, {\cal G}_0\, \lambda^3 -\frac{32 \sqrt{2} }{15 \mu _1^{1/{12}}} q_6\, {\cal G}_1 \lambda^{7/2} + \ldots  \een
\ben \label{11dgvtheta} \frac{g_{v\Theta_i}}{f}  = - \frac{g_{X\Theta_i}}{g} = - \frac{g_{Y\Theta_i}}{h}  =  \frac{256 \sqrt{2}}{135 \mu _1^{5/12}} \, \partial_{\Theta_i} {\cal G}_1\lambda^{9/2}  + \ldots  \een
\be \label{11dgttii} \frac{g_{\Theta_i\Theta_i}}{F_i(\Theta_1, \ldots \Theta_7)}  = \mu_1^{1/3} + \frac{8 }{3 \mu _1^{1/6}} {\cal G}_0\,\lambda^3+ \frac{8 \sqrt{2}}{105 \mu _1^{1/{12}}} \left( 3\,\Delta^{(ii)} + 32\right)({\cal G}_1) \lambda^{7/2}  + \ldots \ee
\be \label{11dgttij} g_{\Theta_i \Theta_j} =  \frac{8 \sqrt{2}}{35 \,\mu_1^{1/12}} \Delta^{(ij)} ({\cal G}_1)\,\lambda^{7/2} + \frac{1}{3}\Delta^{(ij)} ({\cal G}_2)\,\lambda^8 + \ldots \ee
where $z_1$ - $z_3$ and $q_1$ - $q_6$ are the following smooth functions:
\begin{eqnarray} \label{b2}
q_1(v, X, Y) & \equiv &  \partial_X f - \partial_v g, \qquad \qquad    q_2(v, X, Y)  \equiv   -\partial_v f \, \partial_X f+ \partial_v g \, \partial_X g + \partial_v h \, \partial_X h   \nonumber \\
q_3(v, X, Y) & \equiv &  \partial_Y f - \partial_v h, \qquad \qquad q_4(v, X, Y)  \equiv   -\partial_v f \, \partial_Y f+ \partial_v g \, \partial_Y g + \partial_v h \, \partial_Y h \nonumber \\
q_5(v, X, Y) & \equiv & - \left(\partial_X  h + \partial_Y g\right), \qquad  q_6(v, X, Y)  \equiv   -\partial_Y f \, \partial_X f+ \partial_Y g \, \partial_X g + \partial_Y h \, \partial_X h,  \nonumber \\
z_1(v, X, Y) & \equiv &  -\left(\partial_v f \right)^2+\left(\partial_v g \right)^2+\left(\partial_v h \right)^2,  \quad z_2(v, X, Y) \equiv   -\left(\partial_X f \right)^2+\left(\partial_X g \right)^2+\left(\partial_X h \right)^2,  \nonumber \\
z_3(v, X, Y) & \equiv &  -\left(\partial_Y f \right)^2+\left(\partial_Y g \right)^2+\left(\partial_Y h \right)^2,
\end{eqnarray}
the $F_i(\Theta_1, \ldots \Theta_7)$ are defined in \eqref{11dFi} and the $\Delta^{(ij)}$'s are the $28$ second order differential operators given in \eqref{5dDeltao} with the restriction that $1 < i,j < 7$.

The components $g_{vv}$, $g_{XX}$, $g_{YY}$, $g_{vX}$, $g_{vY}$ and $g_{XY}$ are $\mathcal{C}^3$ functions even in the three center solution \cite{Gowdigere:2014aca}. In going from three to four to $\ldots$ to the most generic solution, we expect these components to follow \textbf{(P3)} and be $\mathcal{C}^3$. Clearly the result in \eqref{11dgvv} confirms this expectation.  The components $g_{v\Theta_i}$, $g_{X\Theta_i}$ and $g_{Y\Theta_i}$ for $i=1,2$ are $\mathcal{C}^4$ functions in the three center solution \cite{Gowdigere:2014aca}. In going from three to four to $\ldots$ to the most generic solution, we expect these components to follow \textbf{(P3)} and be $\mathcal{C}^4$. The result in \eqref{11dgvtheta} confirms this expectation. The components $g_{v\Theta_i}$, $g_{X\Theta_i}$ and $g_{Y\Theta_i}$ for $i >2$ are $\mathcal{C}^\infty$ functions in the three center solution \cite{Gowdigere:2014aca}. In going from three to four to $\ldots$ $\infty$ at some stage we expect \textbf{(P2)} and hence they should be at least $\mathcal{C}^3$. The result in \eqref{11dgvtheta} confirms this expectation. Before discussing the $g_{\Theta_i \Theta_j}$ components we note that, just like for the black hole case, we have 
\be \label{11dkernel} \Delta^{(ij)}({\cal G}_1) = 0, \qquad \text{for all $k$-center solutions.}\ee
The diagonal components $g_{\Theta_i \Theta_i}$  are $\mathcal{C}^3$ functions in the three center solution \cite{Gowdigere:2014aca}. In going from three to four to $\ldots$ to the most generic solution, we expect these components to follow \textbf{(P3)} and be $\mathcal{C}^3$. The result in \eqref{11dgttii} together with \eqref{11dkernel} confirms this expectation. The off-diagonal components $g_{\Theta_i \Theta_j}$ for $1 \leq i,j \leq 2$  are $\mathcal{C}^4$ functions in the three center solution \cite{Gowdigere:2014aca}. In going from three to four to $\ldots$ to the most generic solution, we expect these components to follow \textbf{(P3)} and be $\mathcal{C}^4$. The result in \eqref{11dgttij} together with \eqref{11dkernel} confirms this expectation. The off-diagonal components $g_{\Theta_i \Theta_j}$ for $ i,j > 2$  are $\mathcal{C}^\infty$ functions in the three center solution \cite{Gowdigere:2014aca}. In going from three to four to $\ldots$ to the most generic solution, we expect these components to follow \textbf{(P2)} at some stage and be at least $\mathcal{C}^3$. The result in \eqref{11dgttij} together with \eqref{11dkernel} confirms this expectation.

The non-trivial components of the tensor gauge field $C$ in horizon co-ordinates are given by:
\be\label{CvXY} 
C_{vXY} = -\frac{\mu_1^{1/6}}{2}u_2\,\lambda -\frac{2}{\mu_1^{1/6}}u_1\,\lambda ^2 -\frac{8}{\mu_1^{1/2}}\lambda ^3+ \frac{35}{6\mu_1^{1/3}}u_2\,{\cal G}_0\,\lambda ^4 + \frac{736 \sqrt{2}}{135 \mu_1^{1/4}}u_2\, {\cal G}_1\, \lambda ^{9/2}  + \ldots
\ee
\ben \label{CvXl} 
C_{vX\lambda} &=& \frac{\mu_1^{1/2}}{8}u_4 \,\lambda^{-1} -\frac{\mu_1^{1/6}}{2} u_3 - \frac{2}{\mu_1^{1/6}} h\, \lambda - \frac{7}{8}  u_4\, {\cal G}_0\,\lambda ^2 -\frac{104\sqrt{2}\,\mu_1^{1/{12}}}{135} u_4\,  {\cal G}_1\lambda ^{5/2} + \ldots \nonumber \\
 \label{CvYl} 
C_{vY\lambda} &=& \frac{\mu_1^{1/2}}{8}u_6 \,\lambda^{-1} -\frac{\mu_1^{1/6}}{2} u_5 + \frac{2}{\mu_1^{1/6}} g\,\lambda - \frac{7}{8}  u_6\, {\cal G}_0\,\lambda ^2 -\frac{104\sqrt{2}\,\mu_1^{1/{12}}}{135} u_6\,  {\cal G}_1\lambda ^{5/2} + \ldots \nonumber \\
\label{CXYl}
C_{XY\lambda} &=& \frac{\mu_1^{1/2}}{8}u_8 \,\lambda^{-1} -\frac{\mu_1^{1/6}}{2} u_7 - \frac{2}{\mu_1^{1/6}} f\, \lambda - \frac{7}{8}  u_8\, {\cal G}_0\,\lambda ^2 -\frac{104\sqrt{2}\,\mu_1^{1/{12}}}{135} u_8\,  {\cal G}_1\lambda ^{5/2} + \ldots \een

\ben \label{CvXt} 
C_{vX\Theta_i} &=& \frac{32\sqrt{2}\mu_1^{1/{12}}}{135} u_4\, \partial_{\Theta_i}{\cal G}_1\, \lambda ^{7/2} + \frac{2\mu_1^{1/6}}{5}u_4\,  \partial_{\Theta_i} {\cal G}_2\, \lambda ^4 + \ldots \nonumber \\
\label{CvYt} 
C_{vY\Theta_i} &=& \frac{32\sqrt{2}\mu_1^{1/{12}}}{135} u_6\, \partial_{\Theta_i}{\cal G}_1\, \lambda ^{7/2} + \frac{2\mu_1^{1/6}}{5}u_6\,  \partial_{\Theta_i} {\cal G}_2\, \lambda ^4 + \ldots \nonumber \\
\label{CXYt} 
C_{XY\Theta_i} &=& \frac{32\sqrt{2}\mu_1^{1/{12}}}{135} u_8\, \partial_{\Theta_i}{\cal G}_1\, \lambda ^{7/2} + \frac{2\mu_1^{1/6}}{5}u_8\,  \partial_{\Theta_i} {\cal G}_2\, \lambda ^4 + \ldots 
\een
where
\begin{eqnarray} \label{M2b3}
u_1(v, X, Y) & \equiv & \partial_{Y}h + \partial_v f + \partial_X g, \quad
u_3(v, X, Y)  \equiv  h \, \partial_v f - f \, \partial_v h + h \, \partial_X g -g \, \partial_X h, \nonumber \\
u_5(v, X, Y) & \equiv & f \, \partial_v g - g \, \partial_v f + h \, \partial_{Y} g -g \, \partial_{Y} h, \quad
u_7(v, X, Y)  \equiv  f \, \partial_{Y} h - h \, \partial_{Y} f + f \, \partial_X g -g \, \partial_X f, \nonumber \\
u_2(v, X, Y) & \equiv & \left(\partial_v f \, \partial_{Y}h - \partial_{Y}f \, \partial_v h \right) + \left(\partial_X g \, \partial_{Y}h - \partial_{Y}g \, \partial_X h \right)  + \left(\partial_v f \, \partial_{X}g - \partial_{X}f \, \partial_v g \right), \nonumber 
 \een
\ben 
u_4(v, X, Y) & \equiv & f  \left(\partial_v h \, \partial_X g - \partial_X h \, \partial_v g \right) + g \left(\partial_v f \, \partial_X h - \partial_X f \, \partial_v h \right) + h \left(\partial_v g \, \partial_X f - \partial_X g \, \partial_v f \right), \nonumber \\
u_6(v, X, Y) & \equiv & f  \left(\partial_v h \, \partial_{Y} g - \partial_{Y} h \, \partial_v g \right) + g  \left(\partial_v f \, \partial_{Y} h - \partial_{Y} f \, \partial_v h \right)  + h \left(\partial_v g \, \partial_{Y} f - \partial_{Y} g \, \partial_v f \right), \nonumber \\
u_8(v, X, Y) & \equiv & f  \left(\partial_X h \, \partial_{Y} g - \partial_{Y} h \, \partial_X g \right) + g  \left(\partial_X f \, \partial_{Y} h - \partial_{Y} f \, \partial_X h \right) + h  \left(\partial_X g \, \partial_{Y} f - \partial_{Y} g \, \partial_X f \right). \nonumber
\end{eqnarray}
The components $C_{vXY}$is $\mathcal{C}^4$ functions  in the three center solution \cite{Gowdigere:2014aca}. In going from three to four to $\ldots$ to the most generic solution, we expect these components to follow \textbf{(P3)} and be $\mathcal{C}^4$. The result in \eqref{CvXY} confirms this expectation.  The components $C_{vX\lambda}$, $C_{vY\lambda}$ and $C_{XY\lambda}$ are $\mathcal{C}^2$ functions  in the three center solution \cite{Gowdigere:2014aca}. In going from three to four to $\ldots$ to the most generic solution, we expect these components to follow \textbf{(P3)} and be $\mathcal{C}^2$. The result in  \eqref{CvXl} confirms this expectation.  The components $C_{vX\Theta_i}$, $C_{vY\Theta_i}$ and $C_{XY\Theta_i}$ for $i=1,2$ are $\mathcal{C}^3$ functions  in the three center solution \cite{Gowdigere:2014aca}. In going from three to four to $\ldots$ to the most generic solution, we expect these components to follow \textbf{(P3)} and be $\mathcal{C}^3$. The result in  \eqref{CvXt} confirms this expectation.  The components $C_{vX\Theta_i}$, $C_{vY\Theta_i}$ and $C_{XY\Theta_i}$ for $i > 2$ are $\mathcal{C}^\infty$ functions  in the three center solution \cite{Gowdigere:2014aca}. In going from three to four to $\ldots$ to the most generic solution, we expect these components to follow \textbf{(P2)} at some stage and be at least  $\mathcal{C}^2$. The result in  \eqref{CvXt} confirms this expectation. 

Here again, we are able to see that the surmise we made for the $k+1$-center horizon smoothness to the $k$-center horizon smoothness and the consequent expectations are all realized in reality. We have thus shown  that \emph{the horizon smoothness is identical for all $k$-center solutions including the most generic solution, the $\infty$-center solution}.  Not only is the horizon smoothness identical for all $k$-center solutions, we have seen that even an  individual component (whenever it has a finite degree of differentiability) has identical series expansions and hence identical degree of differentiability for all $k$-center solutions (for which it has a finite degree of differentiability). Again we gather the underlying reasons behind this:
\begin{itemize}
\item \textbf{(R1)} The boundary conditions that determine the series ansatze for $r(\lambda)$, $\theta_i(\lambda)$ are identical for all $k$-center solutions. 
\item \textbf{(R2)} In the solution to the geodesic equations, the first $7$  coefficients in the series expansion for $r(\lambda)$,  $c_1, c_2, \ldots c_{7}$ are constant functions as opposed to the a priori possibility that they can be functions of the Gaussian null co-ordinates    $\Theta_i$, for all $k$-center solutions.
\item \textbf{(R3)} The appearance of a set of second order differential operators $\Delta^{(ij)}$ and the fact that each of the summands appearing in the first generalized Gegenbauer polynomial ${\cal G}_1$ are in the kernel of each of them, which implies $\Delta^{(ij)} ({\cal G}_1) = 0 $ for all $k$-center solutions.
\end{itemize}

\subsubsection{\label{307} Solution in an alternate isotropic co-ordinate system}
In this section, we will consider an alternate isotropic co-ordinate system. Instead of \eqref{spcoordinates}, we will choose the following co-ordinates for the transverse $\mathbf{R}^8$, which basically amounts to choosing an alternate co-ordinate system for the seven sphere. 
\ben
x_1 &=& r  \cos \theta_1  \cos \theta_2  \cos \theta_4, \qquad x_2 =  r  \cos \theta_1  \cos \theta_2  \sin \theta_4, \nonumber \\ 
x_3 &=& r  \cos \theta_1  \sin \theta_2  \cos \theta_5, \qquad x_4 =  r  \cos \theta_1  \sin \theta_2  \sin \theta_5, \nonumber \\ 
x_5 &=& r  \sin \theta_1  \cos \theta_3  \cos \theta_6, \qquad x_6 =  r  \sin \theta_1  \cos \theta_3  \sin \theta_6, \nonumber \\ 
x_7 &=& r  \sin \theta_1  \sin \theta_3  \cos \theta_7, \qquad x_8 =  r  \sin \theta_1  \sin \theta_3  \sin \theta_7, 
\label{a11disotropic}
\een
in which the flat metric takes the form
\begin{multline} \label{a11dflatmetric}ds^2_{\mathbf{R}^{8}} = dr^2 + r^2 d\theta_1^2 + r^2\,\cos^2\theta_1\,d\theta_2^2   +  r^2\sin^2\theta_1\,  d\theta_3^2  + r^2 \cos^2 \theta_1\, \cos^2 \theta_2 \,d\theta_4^2 \\ + r^2 \cos^2 \theta_1\, \sin^2 \theta_2 \,d\theta_5^2 + r^2 \sin^2 \theta_1\, \cos^2 \theta_3 \,d\theta_6^2 + r^2 \sin^2 \theta_1\, \sin^2 \theta_3 \,d\theta_7^2 \end{multline}
The definition of the generalized Gegenbauer polynomials proceeds along the lines of \eqref{fR}-\eqref{ggp1} but now we would have 
\begin{multline} \label{11daggp1} {\cal G}_1 (\theta_1, \ldots \theta_7) = \sum_{i=2}^{\infty} \frac{2 \mu_J}{\|\vec{R}^{(J)} \|^{8}}\,\left[R_1^{(J)}  \cos \theta_1  \cos \theta_2  \cos \theta_4 +R_2^{(J)} \cos \theta_1  \cos \theta_2  \sin \theta_4 \right. \\ \left. +R_3^{(J)} \cos \theta_1  \sin \theta_2  \cos \theta_5 + R_4^{(J)} \cos \theta_1  \sin \theta_2  \sin \theta_5 + R_5^{(J)} \sin \theta_1  \cos \theta_3  \cos \theta_6  \right. \\ \left. + R_6^{(J)} \sin \theta_1  \cos \theta_3  \sin \theta_6  + R_7^{(J)} \sin \theta_1  \sin \theta_3  \cos \theta_7 +  R_8^{(J)} \sin \theta_1  \sin \theta_3  \sin \theta_7\right].
\end{multline}
The harmonic function is the formally the same as before, but with the above defined ${\cal G}_n$'s. To construct the Gaussian null co-ordinate system for the first horizon, we will follow the steps laid on in \ref{304}. The only changes to be made from there are that in the $\theta_i$-geodesic equation \eqref{11dthetaigeod} we now have
\begin{multline} \label{11dFia} F_1(\theta_1, \ldots, \theta_7) = 1, \quad F_2(\theta_1, \ldots, \theta_7)  =  \cos^2 \theta_1, \quad F_3(\theta_1, \ldots, \theta_7)  =  \sin^2 \theta_1 \\ F_4(\theta_1, \ldots, \theta_7) = \cos^2 \theta_1\, \cos^2 \theta_2, \quad  F_5(\theta_1, \ldots, \theta_7) = \cos^2 \theta_1\, \sin^2 \theta_2, \\ F_6(\theta_1, \ldots, \theta_7) = \sin^2 \theta_1\, \cos^2 \theta_3, \quad F_7(\theta_1, \ldots, \theta_7) = \sin^2 \theta_1\,\sin^2 \theta_3, \end{multline}
and the null condition \eqref{11dnull1} is appropriately changed.
Note that the boundary conditions \eqref{11dbound} and the final series expansion ansatz \eqref{11dexpansionansatz} are unchanged. Then we ask the question, what is the minimal number of $c_n$'s and $b^{(i)}_n$'s needed to check for our expectations? Due to unchanged series expansion ansatze and the similar formula for the harmonic function, the conclusion does not change. Hence we again have the reduced goal of solving the geodesic equations only upto the point needed to obtain $c_1 - c_8$ and $b^{(i)}_3 - b^{(i)}_9$ for each $i=1,\ldots,7$.

We solve the null condition first and we find we obtain the solution for $r$ which is still given by \eqref{11drtf} but with the generalized Gegenbauer polynomials defined here \ref{307}. We then solve the $\theta_i$-geodesic equations and we obtain solution given by \eqref{11dthetaitf} but with the $F_i(\Theta_1, \ldots, \Theta_7)$ defined in \eqref{11dFia}.

Having obtained the horizon co-ordinate system, we can proceed to compute the tensor components. All the formulae in \ref{306} go through with  three changes: (i) all the ${\cal G}_n$'s are to be replaced with the ${\cal G}_n$'s defined here in \ref{307}, (ii) the $F_i(\Theta_1, \ldots, \Theta_7)$'s appearing in \eqref{11dgttii} are the ones defined in \eqref{11dFia}, (iii) the $\Delta^{(ij)}$'s appearing in \eqref{11dgttii} and \eqref{11dgttij} are the following set of $28$ second differential operators:

\ben 
\Delta^{(11)} &=& 1+ \frac{\partial^2~~}{\partial \Theta_1^2}  \nonumber \\ 
\Delta^{(22)} &=& 1-\tan \Theta_1 \frac{\partial}{\partial \Theta_1} + \frac{1}{\cos \Theta_1^2} \frac{\partial^2}{\partial \Theta_2^2} \nonumber \\
\Delta^{(33)} &=&1+\cot \Theta_1 \frac{\partial ~~}{\partial \Theta_1}   + \frac{1}{\sin^2 \Theta_1} \frac{\partial^2 ~~}{\partial \Theta_3^2}  \nonumber \\ 
\Delta^{(44)} &=& 1 -\tan \Theta_1 \frac{\partial ~~}{\partial \Theta_1} - \frac{\tan \Theta_2}{\cos^2 \Theta_1} \frac{\partial ~~}{\partial \Theta_2}   +\frac{1}{\cos^2 \Theta_1 \cos^2 \Theta_2 } \frac{\partial^2 ~~}{\partial \Theta_4^2} \nonumber \\
\Delta^{(55)}&=& 1 -\tan\Theta_1\frac{\partial~~}{\partial\Theta_1}+ \frac{\cot\Theta_2}{\cos^2\Theta_1}\frac{\partial~~}{\partial\Theta_2} +  \frac{1}{\cos^2\Theta_1\sin^2\Theta_2}\frac{\partial^2~~}{\partial\Theta_5^2} \nonumber \\
\Delta^{(66)}&=& 1 +\cot\Theta_1\frac{\partial~~}{\partial\Theta_1} - \frac{\tan\Theta_3}{\sin^2\Theta_1}\frac{\partial~~}{\partial\Theta_3}  + \frac{1}{\sin^2\Theta_1\cos^2\Theta_3}\frac{\partial^2~~}{\partial\Theta_6^2} \nonumber \\
\Delta^{(77)}&=&1 +\cot\Theta_1\frac{\partial~~}{\partial\Theta_1} + \frac{\cot\Theta_3}{\sin^2\Theta_1}\frac{\partial~~}{\partial\Theta_3}  + \frac{1}{\sin^2\Theta_1\sin^2\Theta_3}\frac{\partial^2~~}{\partial\Theta_7^2} \nonumber \\
\Delta^{(12)} &=& \frac{\partial^2}{\partial \Theta_1 \partial \Theta_2} + \tan \Theta_1  \frac{\partial}{\partial \Theta_2}, \qquad \Delta^{(13)} =\frac{\partial^2}{\partial \Theta_1 \partial \Theta_3} - \cot \Theta_1  \frac{\partial}{\partial \Theta_3}, \nonumber \\
\Delta^{(14)} &=& \frac{\partial^2}{\partial \Theta_1 \partial \Theta_4} + \tan \Theta_1  \frac{\partial}{\partial \Theta_4}, \qquad \Delta^{(15)} =\frac{\partial^2}{\partial \Theta_1 \partial \Theta_5} + \tan \Theta_1  \frac{\partial}{\partial \Theta_5}, \nonumber \\
\Delta^{(16)} &=&\frac{\partial^2}{\partial \Theta_1 \partial \Theta_6} - \cot \Theta_1  \frac{\partial}{\partial \Theta_6}, \qquad \Delta^{(17)} =\frac{\partial^2}{\partial \Theta_1 \partial \Theta_7} - \cot \Theta_1  \frac{\partial}{\partial \Theta_7}, \nonumber \\ 
\Delta^{(23)} &=& \frac{\partial^2}{\partial \Theta_1 \partial \Theta_2} , ~~~~~\qquad \qquad \qquad \Delta^{(24)} =\frac{\partial^2}{\partial \Theta_2 \partial \Theta_4} + \tan \Theta_2  \frac{\partial}{\partial \Theta_4}, \nonumber \\ 
\Delta^{(25)} &=& \frac{\partial^2}{\partial \Theta_1 \partial \Theta_5} - \cot \Theta_2  \frac{\partial}{\partial \Theta_5}, \qquad \Delta^{(26)} =\frac{\partial^2}{\partial \Theta_2 \partial \Theta_6}, \nonumber \\ \vdots \nonumber  \een
\ben \vdots \nonumber \\
\Delta^{(27)} &=& \frac{\partial^2}{\partial \Theta_2 \partial \Theta_7} , \qquad \Delta^{(34)} =\frac{\partial^2}{\partial \Theta_3 \partial \Theta_4}, \qquad \Delta^{(35)} = \frac{\partial^2}{\partial \Theta_3 \partial \Theta_5},   \nonumber \\  
\Delta^{(36)} &=&\frac{\partial^2}{\partial \Theta_3 \partial \Theta_6} + \tan \Theta_3  \frac{\partial}{\partial \Theta_6}, \qquad  \Delta^{(37)} = \frac{\partial^2}{\partial \Theta_3 \partial \Theta_7} - \cot \Theta_3  \frac{\partial}{\partial \Theta_7}, \nonumber \\ 
\Delta^{(45)} &=&\frac{\partial^2}{\partial \Theta_4 \partial \Theta_5} , \qquad \Delta^{(46)} =\frac{\partial^2}{\partial \Theta_4 \partial \Theta_6}, \qquad   \Delta^{(47)} = \frac{\partial^2}{\partial \Theta_4 \partial \Theta_7}, \nonumber \\  
 \Delta^{(56)} &=& \frac{\partial^2}{\partial \Theta_5 \partial \Theta_6}, \qquad \Delta^{(57)} = \frac{\partial^2}{\partial \Theta_5 \partial \Theta_7}, \qquad \Delta^{(67)} = \frac{\partial^2}{\partial \Theta_6 \partial \Theta_7}.
 \een
The remarkable thing is that each of the $8$ functions of the angles that appear as summands in \eqref{11daggp1} (with the $\theta_i$'s replaced by the $\Theta_i$'s) is in the kernel of each of the above $28$ differential operators. What this means is that we still have the result
\be \label{11dkernela} \Delta^{(ij)}({\cal G}_1) = 0, \qquad \text{for all $k$-center solutions.}\ee
 Thus, the result of the computations in this alternate isotropic co-ordinate system \eqref{a11disotropic} is also that: the horizon smoothness is identical for all $k$-center solutions including the most generic solution, the $\infty$-center solution and the degree of differentiability of individual tensor components (when it is finite) is identical for all $k$-center solutions. Again the underlying reasons are the same as the \textbf{(R1)}, \textbf{(R2)} and \textbf{(R3)} given in \ref{306}; but with a different set of $\Delta^{(ij)}$'s  in \textbf{(R3)}.

\section{\label{4} Conclusion and Outlook}
The results of this paper is an end point to a certain line of investigation. The results for the smoothness of the horizon for two center solutions were obtained in \cite{Candlish1} for the black holes and by us in \cite{Gowdigere:2012kq}. Motivated by the fact that there is perhaps some connection between the reduced horizon smoothness of multi center solutions and their reduced symmetry, the investigations in \cite{Gowdigere:2014aca} for three center solutions were taken up and with the aid of  the tool of generalized Gegenbauer polynomials the results were obtained. Even there, after it was shown that the three center and two center horizons have the same smoothness, it became clear that there is no truth to the connection mentioned above. There certainly was no expectation that the less symmetric $k$-center solutions should be less smooth.  But, still the question remained as to what is the smoothness of the horizon in these $k$-center solutions. The results of \cite{Gowdigere:2014aca} and the lessons drawn from them, seem to suggest that all $k$-center solutions have identical horizon smoothness. Still, the task of verifying this, seemed formidable, essentially due to the absence of transverse spatial isometries.  In this paper, we have just performed this computation and verified the expectations that were coming from the investigations in \cite{Gowdigere:2014aca}.

Now that these lengthy calculations have been done, tabulated and reported, one would like to ask the question if there is an easier way to obtain these results.  One easier way could perhaps be to somehow work with the cartesian co-ordinates in the transverse space. The fact that the series expansions for the $g_{\Theta_i \Theta_j}$ seem to be all so similar (see \eqref{5dgii} and \eqref{5dgij}) may be significant. Perhaps the $\Delta^{(ij)}$'s have a simple form in terms of the cartesian co-ordinates. Another easier way  to obtain these results could perhaps be based on the fact that in a particular setup the two center solution and the most generic solution are identical except for the difference in some constants. And perhaps there is a way to show that 
the final answers are independent of these constants. 

A natural question which has been asked before is that of the possibility of making these horizons smooth by considering the multi center solutions as solutions to appropriate higher derivative theories.  For the black hole two center solution, this was already attempted in \cite{Candlish2} for a class of higher derivative terms.  From the results of our work,  we can expect that,  if one were to succeed in analyzing the collinear solution, which is considerably easier due to its many spatial isometries, and show that the horizon is smooth in a certain higher derivative theory, the horizons in the most generic multi center solution will also be smooth.

Finally, we are led to the question of the significance, if any, for M-theory physics, of the result of this paper  that the metric is  ${\cal C}^3$ and the tensor gauge field ${\cal C}^2$ at the multi $M2$ horizon. For example, via the AdS-CFT correspondence, does it have some implication for appropriate correlators in the dual three dimensional field theories? We will leave these investigations for the future. 

\begin{center}
\textbf{Acknowledgments}
\end{center}
I thank Yogesh K Srivastava for many discussions on this subject and for collaborations in \cite{Gowdigere:2012kq} and \cite{Gowdigere:2014aca}. I thank Debashish Goshal, Ashoke Sen and other participants of the National Strings Meet held at IIT-Kharagpur during December 2013 for feedback and comments on this work when it was presented there. I  thank the very friendly staff at the various Cafe Coffee Day outlets in Bhubaneshwar, where quite a bit of this work was done, for their warm hospitality.

\appendix
\section{\label{A}Series expansions for tensor components prior to solving the geodesic equations}

Prior to solving the geodesic equations and constructing the Gaussian null co-ordinate system for the horizon in the most generic $d=5$ black hole solution, in \ref{211}, we asked the question: What is the minimal number of the coefficients $c_n$'s and the $b^{(i)}_n$'s  needed to check for the expectations one has for  the horizon smoothness of the most generic solution? Here we collect all the formule needed to answer this question, obtained by plugging in the series expansion ansatze \eqref{expansionansatz} into the tensor transformation laws.  In the following, we should note that the coefficeints are functions: $c_n(\Theta_1, \Theta_2, \Theta_3), b^{(i)}_n(\Theta_1, \Theta_2, \Theta_3).$ 

We will first give the formula for the $T(\lambda, \Theta_1, \Theta_2, \Theta_3)$, obtained using \eqref{expansionansatz} and  \eqref{228}:
\begin{multline} \label{T}
T(\lambda, \Theta_1, \Theta_2, \Theta_3) = -\frac{\mu _1^2}{c_1^4}\,\lambda^{-1} + \frac{8 \mu _1^2\,c_2 }{c_1^5}\,\lambda^{-1/2} + \frac{2 \mu _1}{c_1^6} \left[ c_1^4 {\cal G}_0 - 2 \mu_1 c_3 c_1 +5 \mu_1 c_2^2  \right] \log \lambda + \frac{4 \mu _1}{c_1^7} \left[ -2 c_2 c_1^4 {\cal G}_0 \right. \\ \left. +c_1^6 {\cal G}_1 - 2 \mu_1\, c_4 c_1^2  + 10 \mu_1\, c_3 c_2 c_1  - 10 \mu _1\, c_2^3 \right] \lambda^{1/2} + \frac{1}{c_1^8} \left[ c_1^8 {\cal G}_0^2 + 6 \mu _1\, c_2^2 c_1^4  {\cal G}_0 - 4  \mu_1 \,c_3 c_1^5 {\cal G}_0 - 2 \mu_1\,c_2 c_1^6 {\cal G}_1 \right. \\ \left. + 2 \mu_1\, c_1^8 {\cal G}_2    -4 \mu _1^2\, c_5 c_1^3 +10 \mu _1^2\, c_3^2 c_1^2 + 20\mu _1^2\, c_4 c_2 c_1^2 - 60 \mu _1^2\, c_3 c_2^2 c_1 + 35\mu _1^2\, c_2^4 \right] \lambda + \frac{4}{3 c_1^9} \left[ c_1^{10} {\cal G}_1 {\cal G}_0  -2\mu_1\, c_4 c_1^6 {\cal G}_0  \right. \\ \left. + 6\mu_1\, c_3 c_2 c_1^5 {\cal G}_0 - 4\mu_1\, c_2^3 c_1^4 {\cal G}_0  - \mu_1\, c_3 c_1^7 {\cal G}_1 + \mu_1 \,c_2^2\, c_1^6 {\cal G}_1 + \mu_1\,c_1^{10} {\cal G}_3 - 28 \mu_1^2\, c_2^5 + 70 \mu_1^2\,  c_3 c_2^3 c_1 - 30\mu_1^2\,  c_4 c_2^2 c_1^2 \right. \\ \left.  - 30 \mu_1^2\,  c_3^2 c_2 c_1^2  + 10  \mu_1^2\, c_5 c_2 c_1^3 + 10 \mu_1^2\,  c_4 c_3 c_1^3 - 2 \mu_1^2\,  c_6 c_1^4\right] \lambda^{3/2} + \ldots 
\end{multline}
We will also need
\begin{multline} \label{H} 
H(\lambda, \Theta_1, \Theta_2, \Theta_3) =  \frac{\mu _1}{c_1^2} \lambda^{-1} -\frac{2 \mu_1\, c_2 }{c_1^3} \lambda^{-1/2} + \frac{1}{c_1^4} \left[ c_1^4 {\cal G}_0 - 2 \mu_1\,c_3 c_1 + 3 \mu_1\, c_2^2 \right] + \frac{1}{c_1^5} \left[c_1^6 {\cal G}_1 - 2 \mu_1\, c_4 c_1^2  \right. \\ \left.+ 6 \mu_1\, c_3 c_2 c_1 - 4 \mu_1 \, c_2^3  \right] \lambda^{1/2} + \frac{1}{c_1^6} \left[ c_1^8 {\cal G}_2 + c_2 c_1^6 {\cal G}_1 - 2 \mu _1\, c_5 c_1^3 +3 \mu _1 \, c_3^2 c_1^2 + 6 \mu _1\, c_4 c_2 c_1^2 - 12 \mu _1\, c_3 c_2^2  c_1  \right. \\ \left.  + 5\mu _1 \, c_2^4 \right] \lambda + \frac{1}{c_1^7} \left[ c_1^{10} {\cal G}_3 + 2 c_2 c_1^8 {\cal G}_2 + c_3 c_1^7 {\cal G}_1 - 2 \mu_1\,c_6 c_1^4 + 6 \mu_1\,  c_4 c_3 c_1^3 + 6 \mu_1\, c_5 c_2 c_1^3 - 12 \mu_1\,  c_3^2 c_2 c_1^2 \right. \\ \left. - 12 \mu_1\, c_4 c_2^2 c_1^2 + 20 \mu_1 \,  c_2^3 c_3 c_1 - 6 \mu_1\, c_2^5 \right] \lambda^{3/2} + \ldots,\end{multline}
which is obtained using \eqref{expansionansatz} in \eqref{5dharm}.

The series expansions for the metric components in the Gaussian null co-ordinate system are obtained using \eqref{expansionansatz} in the tensor transformation laws:
\begin{multline} \label{5dgvvbefore} g_{vv} = - \frac{1}{H^2} \\ \qquad = -\frac{c_1^4}{\mu _1^2}\,\lambda^2  -\frac{4 c_2 c_1^3}{\mu _1^2} \lambda^{5/2}   + \frac{2}{\mu_1^3} \left[ c_1^6 {\cal G}_0-2 c_3 c_1^3 \mu _1-3 c_2^2 c_1^2 \mu _1\right] \lambda^3   + \frac{2}{\mu_1^3} \left[ 6 c_2 c_1^5 {\cal G}_0+c_1^7 {\cal G}_1-2 c_4 c_1^3 \mu _1 \right. \\ \left.  -6 c_2 c_3 c_1^2 \mu _1-2 c_2^3 c_1 \mu _1\right] \lambda^{7/2} + \ldots\end{multline}
From above,  we can conclude that to examine $g_{vv}$ up to order $\lambda^{5/2}$ we would need the expressions for $c_1$ and $c_2$ only; to compute till order $\lambda^{7/2}$ we would need the expressions for $c_1, c_2, c_3, c_4$. 
\be g_{v\Theta_i} =  \frac{\partial_{\Theta_i}T}{H^2}.\ee
 Using \eqref{T} and \eqref{5dgvvbefore}, we obtain
\begin{multline} \label{5dgvtbefore} g_{v\Theta_i} = \frac{4\, \partial_{ \Theta_i}c_1}{c_1} \lambda - \frac{8}{c_1^2} \left[ 3\, c_ 2\, \partial_{\Theta_i}c_ 1 - c_ 1\, \partial_{\Theta_i}c_ 2\right] \lambda^{3/2} - \frac{8}{\mu _1 c_1^3 } \left[ c_1^4\,  \partial_{\Theta_i}c_1\, {\cal G}_0 - 2 \mu_1\, c_3 \,c_1\,  \partial_{\Theta_i}c_1 - 4 \mu_1 \, c_2\,  c_1\,  \partial_{\Theta_i}c_2 \right. \\ \left. + 17 \mu_1 \, c_2^2 \, \partial_{\Theta_i}c_1 \right] \lambda^2 + \frac{4}{\mu _1 c_1^4 } \left[ c_1^7 {\cal G}_1 - 3 c_1^6\, \partial_{\Theta_i}c_1  \,{\cal G}_1 - 6 c_1^5 \, \partial_{\Theta_i}c_2 \, {\cal G}_0 + 14 c_2\, c_1^4 \, \partial_{\Theta_i}c_1 \, {\cal G}_0 - 2 \mu _1\,  c_1^3  \, \partial_{\Theta_i}c_4 \right. \\ \left. + 14 \mu_1 \,c_4\, c_1^2\,  \partial_{\Theta_i}c_1 + 18 \mu_1 \,c_3\, c_1^2\,  \partial_{\Theta_i}c_2 + 10 \mu_1 \, c_2\, c_1^2\,  \partial_{\Theta_i}c_3 - 88 \mu_1\, c_3\, c_2\, c_1  \,\partial_{\Theta_i}c_1 - 18 \mu_1\,  c_2^2\, c_1\,  \partial_{\Theta_i}c_2 \right. \\ \left. + 14 \mu_1 \,c_2^3 \, \partial_{\Theta_i}c_1 \right] \lambda^{5/2} + \ldots - \frac{4}{\mu _1 \,c_1^3} \left[ c_1^4\,  \partial_{\Theta_i}c_1\,{\cal G}_0 + \mu_1 c_1^2\, \partial_{\Theta_i}c_3 - 5 \mu _1 \,c_3\, c_1  \partial_{\Theta_i}c_1 - 5 \mu_1 \, c_2\, c_1\, \partial_{\Theta_i}c_2 \right. \\ \left. + 15 \mu_1\, c_2^2 \, \partial_{\Theta_i}c_1 \right]  \log \lambda \left[ \lambda^2 + \frac{4 c_2}{c_1}\lambda^{5/2} + \ldots  \right]\end{multline}
From above, we can conclude that to examine $g_{v\Theta_i}$ up to order $\lambda^{3/2}$ we would need the expressions for $c_1$ and $c_2$ only; to compute till order $\lambda^{5/2}$ we would need the expressions for $c_1, c_2, c_3, c_4$. The rest of the metric components are as follows:

\begin{multline} \label{5dgtt11befor} g_{\Theta_1\Theta_1} = -   \frac{(\partial_{\Theta_1}T)^2}{H^2} + (\partial_{\Theta_1}r)^2 H + (\partial_{\Theta_1}\theta_1)^2 H \,r^2 + (\partial_{\Theta_1}\theta_2)^2 H \,r^2\,\sin \theta_1^2 + (\partial_{\Theta_1}\theta_3)^2 H \,r^2\,\sin \theta_1^2\,\sin \theta_2^2 \\ = -   \frac{(\partial_{\Theta_1}T)^2}{H^2} + (\partial_{\Theta_1}r)^2 H + \mu_1 + c_1^2 {\cal G}_0\, \lambda + \left[  c_1^3 {\cal G}_1   + 2 \mu_1 \, c_2 c_1 {\cal G}_0+ 2 \mu _1\, \partial_{\Theta_1}b^{(1)}_3 \right] \lambda^{3/2}   \\ + \left[  c_1^4 {\cal G}_2+3 c_2 c_1^2 {\cal G}_1 + 2 c_3 c_1 {\cal G}_0 + c_2^2 {\cal G}_0 + 2  \mu _1\, \partial_{\Theta_1}b^{(1)}_4\right] \lambda^2 + \left[c_1^5 {\cal G}_3 +4 c_2 c_1^3 {\cal G}_2 +3 c_3 c_1^2 {\cal G}_1 + 3 c_2^2 c_1 {\cal G}_1 \right. \\ \left. +2 c_4 c_1 {\cal G}_0 + 2 c_3 c_2 {\cal G}_0 + 2 c_1^2 \,\partial_{\Theta_1}b^{(1)}_3 \, {\cal G}_0 + 2  \mu _1\,\partial_{\Theta_1} b^{(1)}_5\right] \lambda^{5/2} + \ldots \end{multline}

\begin{multline}  \label{5dgtt22before} g_{\Theta_2\Theta_2} = - \frac{(\partial_{\Theta_2}T)^2}{H^2} + (\partial_{\Theta_2}r)^2 H + (\partial_{\Theta_2}\theta_1)^2 H \,r^2 + (\partial_{\Theta_2}\theta_2)^2 H \,r^2\,\sin \theta_1^2 + (\partial_{\Theta_2}\theta_3)^2 H \,r^2\,\sin \theta_1^2\,\sin \theta_2^2  \\ = -   \frac{(\partial_{\Theta_2}T)^2}{H^2} + (\partial_{\Theta_2}r)^2 H + \mu_1\sin^2 \Theta_1 + c_1^2 {\cal G}_0 \sin^2 \Theta_1\, \lambda + \left[  c_ 1^3 {\cal G}_ 1 + 2 c_ 2 c_ 1 {\cal G}_0  + 2 \mu _ 1 \,b^{(1)}_ 3   \cot \Theta  \right. \\ \left.  + 2 \mu _ 1  \,\partial_{\Theta_2}b^{(2)}_ 3\right] \sin^2 \Theta_1\, \lambda^{3/2}  + \left[ c_ 1^4 {\cal G}_ 2 + 3 c_ 2 c_ 1^2 {\cal G}_ 1+2 c_ 3 c_ 1 {\cal G}_0 + c_ 2^2 {\cal G}_0 + 2 \mu _ 1\, b^{(1)}_4 \cot \Theta  + 2 \mu_1\,\partial_{\Theta_2}b^{(2)}_ 4  \right] \sin^2 \Theta_1\, \lambda^2 \\ + \left[ c_ 1^5 {\cal G}_ 3 + 4 c_ 2 c_ 1^3 {\cal G}_ 2 + 3 c_ 3 c_ 1^2  {\cal G}_ 1 + 3 c_ 2^2 c_ 1 {\cal G}_ 1 + 2 c_ 1^2 b^{(2)}_ 3 {\cal G}_0 + 2 c_ 4 c_ 1 {\cal G}_0 + 2 c_ 2 c_ 3 {\cal G}_0 + 2  c_ 1^2 b^{(1)}_ 3 {\cal G}_0 \cot \Theta + 2 \mu _ 1\,b^{(1)}_ 5  \cot \Theta  \right. \\ \left.  + 2 \mu _ 1 \, \partial_{\Theta_2}b^{(2)}_ 5 \right]\sin^2 \Theta_1\, \lambda^{5/2} + \ldots \end{multline}  
 
\begin{multline}  \label{5dgtt33before} g_{\Theta_3\Theta_3} = - \frac{(\partial_{\Theta_3}T)^2}{H^2} + (\partial_{\Theta_3}r)^2 H + (\partial_{\Theta_3}\theta_1)^2 H \,r^2 + (\partial_{\Theta_3}\theta_2)^2 H \,r^2\,\sin \theta_1^2 + (\partial_{\Theta_3}\theta_3)^2 H \,r^2\,\sin \theta_1^2\,\sin \theta_2^2  \\ = -   \frac{(\partial_{\Theta_3}T)^2}{H^2} + (\partial_{\Theta_3}r)^2 H + \mu_1\sin^2 \Theta_1 \sin^2 \Theta_2 + c_1^2 {\cal G}_0 \sin^2 \Theta_1 \sin^2 \Theta_2\, \lambda + \left[2 c_ 2 c_ 1 {\cal G}_0 + c_ 1^3 {\cal G}_ 1 + 2  \mu_1 \,b^{(1)}_ 3 \cot \Theta_1 \right. \\ \left. + 2 \mu_1 \,b^{(2)}_ 3\cot \Theta_2 +  2\mu _ 1 \, \partial_{\Theta_3}b^{(3)}_ 3  \right] \sin^2 \Theta_1 \sin^2 \Theta_2\, \lambda^{3/2}  + \left[ c_ 1^4 {\cal G}_ 2 + 3 c_ 2 c_ 1^2 {\cal G}_ 1 + 2 c_ 3 c_ 1 {\cal G}_0 + c_ 2^2 {\cal G}_0 + 2 \mu _ 1 \,b^{(1)}_ 4  \cot \Theta_1 \right. \\ \left.  + 2 \mu_1\, b^{(2)}_ 4  \cot \Theta_2 +2 \mu _ 1 \, \partial_{\Theta_3}b^{(3)}_ 4\right] \sin^2 \Theta_1 \sin^2 \Theta_2\, \lambda^2   + \left[c_ 1^5 {\cal G}_ 3+4 c_ 2 c_ 1^3 {\cal G}_ 2+3 c_ 3 c_ 1^2 {\cal G}_ 1+3 c_ 2^2 c_ 1 {\cal G}_ 1+2 c_ 4 c_ 1 {\cal G}_ 0 \right. \\ \left. +2 c_ 2 c_ 3 {\cal G}_ 0  +2 b^{(1)}_ 3 c_ 1^2 {\cal G}_ 0 \cot \Theta_1  + 2 \mu _ 1 \, b^{(1)}_ 5    \cot \Theta_1 +2 c_ 1^2 b^{(2)}_ 3 {\cal G}_ 0 \cot \Theta_2 + 2 \mu _ 1 \, b^{(2)}_ 5    \cot \Theta_2 +2 c_ 1^2 {\cal G}_ 0 \partial_{\Theta_3}b^{(3)}_ 3 \right. \\ \left. +2 \mu _ 1 \,\partial_{\Theta_3}b^{(3)}_ 5 \right]\sin^2 \Theta_1 \sin^2 \Theta_2\, \lambda^{5/2} + \ldots \end{multline}

\begin{multline} \label{5dgtt12before} g_{\Theta_1\Theta_2} = - \frac{\partial_{\Theta_1}T\,\partial_{\Theta_2}T}{H^2} + \partial_{\Theta_1}r\,\partial_{\Theta_2}r\, H + \partial_{\Theta_1}\theta_1\,\partial_{\Theta_2}\theta_1\, H \,r^2 + \partial_{\Theta_1}\theta_2\,\partial_{\Theta_2}\theta_2\, H \,r^2\,\sin \theta_1^2  \\ + \partial_{\Theta_1}\theta_3\,\partial_{\Theta_2}\theta_3\,  H \,r^2\,\sin \theta_1^2\,\sin \theta_2^2  = - \frac{\partial_{\Theta_1}T\,\partial_{\Theta_2}T}{H^2} + \partial_{\Theta_1}r\,\partial_{\Theta_2}r\, H  + \left[ \mu_1\,\partial_{\Theta_2}b^{(1)}_ 3 + \mu_1\,\sin^2\Theta_1 \partial_{\Theta_1}b^{(2)}_ 3\right]  \lambda^{3/2} \\ + \left[ \mu_1\,\partial_{\Theta_2}b^{(1)}_ 4 + \mu_1\,\sin^2\Theta_1 \partial_{\Theta_1}b^{(2)}_ 4\right]  \lambda^2   + \left[c_ 1^2 {\cal G}_ 0 \partial_{\Theta_2}b^{(1)}_ 3+\mu _ 1\, \partial_{\Theta_2}b^{(1)}_ 5 + c_ 1^2 {\cal G}_ 0 \sin ^2\Theta_1  \partial_{\Theta_1}b^{(2)}_ 3 \right. \\ \left.+\mu _ 1 \sin ^2\Theta_1 \partial_{\Theta_1} b^{(2)}_ 5\right] \lambda^{5/2} + \ldots \end{multline}  

\begin{multline}  \label{5dgtt13before} g_{\Theta_1\Theta_3} = - \frac{\partial_{\Theta_1}T\,\partial_{\Theta_3}T}{H^2} + \partial_{\Theta_1}r\,\partial_{\Theta_3}r\, H + \partial_{\Theta_1}\theta_1\,\partial_{\Theta_3}\theta_1\, H \,r^2 + \partial_{\Theta_1}\theta_2\,\partial_{\Theta_3}\theta_2\, H \,r^2\,\sin \theta_1^2  \\ + \partial_{\Theta_1}\theta_3\,\partial_{\Theta_3}\theta_3\,  H \,r^2\,\sin \theta_1^2\,\sin \theta_2^2  = - \frac{\partial_{\Theta_1}T\,\partial_{\Theta_3}T}{H^2} + \partial_{\Theta_1}r\,\partial_{\Theta_3}r\, H  + \left[ \mu_1\,\partial_{\Theta_3}b^{(1)}_ 3 + \mu_1\,\sin^2\Theta_1 \sin^2\Theta_2 \, \partial_{\Theta_1}b^{(3)}_ 3\right]  \lambda^{3/2} \\ + \left[ \mu_1\,\partial_{\Theta_3}b^{(1)}_ 4 + \mu_1\,\sin^2\Theta_1 \sin^2\Theta_2 \,\partial_{\Theta_1}b^{(3)}_4\right]  \lambda^2   + \left[c_ 1^2 {\cal G}_ 0 \partial_{\Theta_3}b^{(1)}_ 3+\mu _ 1\, \partial_{\Theta_3}b^{(1)}_ 5 + c_ 1^2 {\cal G}_ 0 \sin ^2\Theta_1 \sin^2\Theta_2 \,  \partial_{\Theta_2}b^{(3)}_ 3 \right. \\ \left. +\mu _ 1 \sin ^2\Theta_1 \sin^2\Theta_2 \, \partial_{\Theta_2} b^{(3)}_ 5\right] \lambda^{5/2} + \ldots \end{multline}

\begin{multline}  \label{5dgtt23befor} g_{\Theta_2\Theta_3} = - \frac{\partial_{\Theta_2}T\,\partial_{\Theta_3}T}{H^2} + \partial_{\Theta_2}r\,\partial_{\Theta_3}r\, H + \partial_{\Theta_2}\theta_1\,\partial_{\Theta_3}\theta_1\, H \,r^2 + \partial_{\Theta_2}\theta_2\,\partial_{\Theta_3}\theta_2\, H \,r^2\,\sin \theta_1^2  \\ + \partial_{\Theta_2}\theta_3\,\partial_{\Theta_3}\theta_3\,  H \,r^2\,\sin \theta_1^2\,\sin \theta_2^2  = - \frac{\partial_{\Theta_2}T\,\partial_{\Theta_3}T}{H^2} + \partial_{\Theta_2}r\,\partial_{\Theta_3}r\, H  + \left[ \mu_1 \sin^2 \Theta_1 \partial_{\Theta_3}b^{(2)}_ 3 \right. \\ \left. + \mu_1\,\sin^2\Theta_1 \sin^2\Theta_2 \, \partial_{\Theta_2}b^{(3)}_ 3\right]  \lambda^{3/2} + \left[\mu_1 \sin^2 \Theta_1 \partial_{\Theta_3}b^{(2)}_ 4 + \mu_1\,\sin^2\Theta_1 \sin^2\Theta_2 \, \partial_{\Theta_2}b^{(3)}_ 4\right]  \lambda^2   + \left[c_ 1^2 {\cal G}_ 0 \sin^2 \Theta_1\,\partial_{\Theta_3}b^{(2)}_ 3 \right. \\ \left. +\mu _ 1\,\sin^2  \Theta_1 \,\partial_{\Theta_3}b^{(2)}_ 5  + c_ 1^2 {\cal G}_ 0 \sin ^2\Theta_1 \sin^2\Theta_2 \,  \partial_{\Theta_1}b^{(3)}_ 3  +\mu _ 1 \sin ^2\Theta_1 \sin^2\Theta_2 \, \partial_{\Theta_1} b^{(3)}_ 5\right] \lambda^{5/2} + \ldots \end{multline}  

The term $- \frac{\partial_{\Theta_i}T\,\partial_{\Theta_j}T}{H^2} + \partial_{\Theta_i}r\,\partial_{\Theta_j}r\, H$ is common to the above six formulae \label{5dgtt11before} \label{5dgtt23before}. Using \eqref{T} and \eqref{H}, we can conclude that to examine this term up to order $\lambda^{3/2}$ we would need the expressions for $c_1$ and $c_2$ only; to compute till order $\lambda^{5/2}$ we would need the expressions for $c_1, c_2, c_3, c_4$. Hence, to examine $g_{\Theta_i \Theta_j}$ up to order $\lambda^{3/2}$,  we conclude that we only need the co-efficients $c_1-c_4$ and $b^{(i)}_3$ for all $i = 1,2,3$. If we need to compute the first non-zero odd order i.e. order five, we would need no more than the coefficients $c_1 - c_6$ and $b^{(i)}_3 - b^{(i)}_5$ for all $i = 1,2,3$.

The series expansions for the gauge field components in the Gaussian null co-ordinate system are obtained using \eqref{expansionansatz} in the tensor transformation laws:

\be \label{5dAbefore} A_\lambda = H, \quad A_v  = - H^{-1}, \quad  A_{\Theta_i} = \frac{\partial_{\Theta_i}T}{H}.\ee
Using \eqref{T} and \eqref{H} we can conclude that to examine $A_{v}$ up to order $\lambda^{3/2}$, $A_\lambda$ up tp order $\lambda^{-1/2}$ and $A_{\Theta_i}$ up to order $\lambda^{1/2}$ we would need the expressions for $c_1-c_4$; to compute till  the next odd order in the $\lambda^{1/2}$-expansion, we would need the expressions for $c_1-c_6$.

\end{document}